\pdfoutput=1

\documentclass[fleqn,usenatbib,usedcolumn]{mnras}

\usepackage{newtxtext,newtxmath}

\usepackage[T1]{fontenc}
\usepackage{ae,aecompl}

\usepackage{graphicx}	
\usepackage{amsmath}	
\usepackage{amssymb}	

\graphicspath{{.}{plots/}}

\title[Modelling a cosmic void]{An alternative approach to
modelling a cosmic void and its effect on the cosmic microwave background}

\author[Do Young Kim et al.]{
Do Young Kim,$^{1,2}$\thanks{dyk25@mrao.cam.ac.uk}
Anthony N. Lasenby,$^{1,2}$\thanks{a.n.lasenby@mrao.cam.ac.uk}
and Michael P. Hobson$^{1}$\thanks{mph@mrao.cam.ac.uk}
\\
$^{1}$Astrophysics Group, Cavendish Laboratory, JJ Thomson Avenue, Cambridge CB3 0HE\\
$^{2}$Kavli Institute for Cosmology, Madingley Road, Cambridge CB3 0HA\\
}

\date{Accepted XXX. Received YYY; in original form ZZZ}

\pubyear{2015}

\volume{{\rm in press}}

\begin{document}
\label{firstpage}
\pagerange{\pageref{firstpage}--\pageref{lastpage}}
\maketitle

\begin{abstract}
We apply our tetrad-based approach for constructing
spherically-symmetric solutions in general relativity to modelling
a void, and compare it with the standard Lema\^{\i}tre--Tolman--Bondi (LTB)
formalism. In particular, we construct models for the void observed in
the direction of Draco in the WISE-2MASS galaxy survey, and a
corresponding cosmic microwave background (CMB) temperature decrement in the Planck data in the same
direction. We find that the present-day density and velocity profiles
of the void are not well constrained by the existing data, so that
void models produced from the two approaches can differ substantially
while remaining broadly consistent with the observations. We highlight the importance of considering the velocity as well as the density profile in constraining voids.
\end{abstract}




\section{Introduction}

It is of interest in cosmology to model non-linear structures
such as clusters and voids, and determine the secondary temperature
anisotropies that they induce in the cosmic microwave background (CMB).
Recent attention has focussed in particular on voids,
which arise naturally in $\Lambda$CDM cosmologies through the
evolution of large scale structure, surrounded by filaments and
clusters in the cosmic web \citep{Colless2001,
  Tegmark2003,Sutter2012}. Indeed, voids are of particular interest
since their distribution is sensitive to the equation of state of dark
energy \citep{Pisani2015,Lavaux2011}.  Moreover, the presence of our
Galaxy within a large local void has been suggested as an alternative
explanation for observations of the acceleration of the universal
expansion, without invoking dark energy \citep{Celerier2012,
  Celerier2012a, Bolejko2010, Bene2010, Kainulainen2009, Marra2007,
  Marra2007a, Alexander2007} although it is likely that only a small part of the observed acceleration could be due to such an effect \citep{Geshnizjani2005, Siegel2005, Zibin2008b}.

Individual clusters and voids are often modelled as spherically-symmetric,
pressureless systems using the Lema\^{\i}tre--Tolman--Bondi (LTB) metric
\citep{Romano2015, Tokutake2016, Finelli2016, Brouzakis2006}. 
The LTB model is usually expressed in comoving coordinates
and thus provides a Lagrangian picture of the fluid evolution. 
Such models can accommodate an arbitrary, usually continuous, density
profile for the central object, but do have some limitations.  For
example, the central object is usually only compensated at infinity,
which can complicate the interpretation of observational effects,
since observers comoving with the cosmological fluid are not in a
region modelled by a homogeneous Friedmann--Robertson--Walker (FRW)
cosmology. In principle, compensation at a finite radius can be
achieved by an appropriate choice of initial radial density and
velocity profiles, but in so doing care must be taken to avoid
subsequent streamline crossing, since the presence of shock fronts
would necessitate the inclusion of pressure to produce a realistic
model. 
Finally, the LTB metric
contains a residual gauge freedom that necessitates the imposition of
arbitrary initial conditions to determine the system evolution.

As a consequence, we have for some time adopted a different,
tetrad-based method for solving the Einstein field equations for
spherically-symmetric systems \citep{Lasenby1998, NLH1, NLH2, NLH3,
  Kim2016a}. Aside from straightforwardly accommodating pressure
(which we will not consider here), the method has no gauge ambiguities
in non-vacuum regions and is expressed in terms of a non-comoving
radial coordinate that results in a Eulerian picture of the fluid
evolution with a clear physical interpretation. Indeed, the gauge
choices employed lead to dynamical equations that are essentially
Newtonian in form. Assuming a pressureless fluid throughout, we have
already applied the method to modelling the evolution of a
finite-size, spherically-symmetric cluster, with continuous radial
density and velocity profiles, that is embedded in an expanding
background universe and compensated so that it does not exert any
gravitational influence on the exterior universe \citep{Lasenby1999,
  Dabrowski1999,Dabrowski2000}.  In our approach, one considers an
initial velocity profile from which the initial density profile is
determined uniquely by the constraints that there are no decaying
modes present and that the density distribution is
compensated. Moreover, this compensation holds at all later times, and
the velocity field evolves in a way that avoids streamline crossing.


In this paper, we apply our approach to modelling voids and calculate their effect on the CMB. As a particular example, we consider the Draco supervoid, for which the present day density and velocity profiles have been estimated by \cite{Finelli2016} (hereinafter FGKPS) from a projected underdensity in the WISE-2MASS galaxy survey and a CMB temperature decrement in the Planck data in the same direction. We consider a number of ways in which a similar void can be produced in our approach, and determine the resulting temperature decrements, with particular focus on the influence of the void velocity profile. We also compare our results with those derived previously using the LTB model. 

The structure of this paper is as follows. In Section~\ref{sec:tetrad}, we introduce the tetrad-based method, and the model we use for spherical perturbations that are consistent with having evolved from primordial fluctuations in the early universe. In Section~\ref{sec:LTB}, we discuss the LTB void model used by \cite{Finelli2016, Mackenzie2017, Marcos-Caballero2016, Zibin2014, Nadathur2014} in their analyses of supervoids. We then compare CMB decrements caused by voids similar to the LTB model used to represent the Draco supervoid by FGKPS in Sections~\ref{sec:mimic1} and \ref{sec:mimic2}. Lastly we present our conclusions in Section~\ref{sec:conc}.



\section{Tetrad-based methodology and void model}
\label{sec:tetrad}

Our tetrad-based approach is summarised in \cite{Kim2016a}.  In a
Riemannian spacetime, the relationship between the coordinate basis
vectors $\mathbf{e}_{\mu}$ and local Lorentz frame vectors
$\hat{\mathbf{e}}_a$ are given by the tetrads, or vierbeins
${e_{a}}^{\mu}$, where the inverse is denoted ${e^{a}}_{\mu}$, such
that $\hat{\mathbf{e}}_a={e_{a}}^{\mu}\mathbf{e}_{\mu}$ and
$\mathbf{e}_{\mu}={e^{a}}_{\mu}\hat{\mathbf{e}}_a$. Assuming spherical
symmetry and a pressureless fluid, one may adopt a gauge in which the
non-zero tetrad components are given by ${e_{0}}^{0}=1$,
${e_0}^{1}=g_2$, ${e_1}^{1}=g_1$, ${e_2}^{2}=1/r$ and
${e_3}^{3}=1/(r\sin\theta)$, where $g_1(r,t)$ and $g_2(r,t)$ are
unknown functions.  Note that dependencies on both $r$ and $t$ will
often be suppressed in the equations presented below, whereas we will
usually make explicit dependency on either $r$ or $t$ alone. The
gauge adopted is called the Newtonian gauge (not to be confused with
that used in perturbation theory) because it allows simple Newtonian
interpretations of the dynamics. In this gauge, the metric
coefficients derived from the tetrad components lead to the
line-element
\begin{equation}
ds^2=\left(1 - \frac{g_2^{2}}{g_1^{2}}\right)\,dt^2+\frac{2g_2}{g_1^{2}}\,dt\,dr-\frac{1}{g_1^{2}}\,dr^2-r^2d\Omega^2.\label{eq:generalmetric}
\end{equation}
The time coordinate $t$ measures the proper time of observers comoving
with the fluid, and the (non-comoving) radial coordinate $r$ labels
spheres of proper area $4\pi r^2$. It is straightforward to show that
$g_2$ is the rate of change of the $r$ coordinate of a fluid particle
(or comoving observer) with respect to its proper time, and so can be
physically interpreted as the fluid 3-velocity.  As demonstrated below,
the physical interpretation of $g_1$ is such that the total energy per
unit mass of a fluid particle is $\frac{1}{2}(g_1^2-1)$ (after
subtraction of the rest-mass energy).

\subsection{Field equations}

For a fluid with density $\rho(r,t)$, the total mass-energy, $M(r,t)$,
contained in a sphere of radius $r$ is given by
\begin{equation}
    \frac{\partial M}{\partial r} = 4 \pi r^2 \rho,
    \label{eqn:dMdr}
\end{equation}
and the mass flow through a sphere of radius $r$ per unit time is given by
\begin{equation}
    \frac{\partial M}{\partial t} = -4 \pi r^2 \rho g_2.
    \label{eqn:massflow}
\end{equation}
As mentioned above, the variable $g_2$ may be interpreted physically
as the fluid 3-velocity in the coordinate frame and defines the integral
curves of the conserved fluid current by
\begin{equation}
\frac{dr}{dt}=g_2.
\label{eqn:vel}
\end{equation}
These integral curves are also matter geodesics, since the fluid is pressureless. For functions of both $r$ and $t$, one can therefore define the comoving derivative
\begin{equation}
    \frac{d}{dt}
= \frac{\partial}{\partial t} + g_2 \frac{\partial}{\partial r},
    \label{eqn:comoving_deriv}
\end{equation}
which determines the rate of change of a quantity along a streamline
with respect to the proper time of a comoving fluid element. From
equations \eqref{eqn:dMdr} and \eqref{eqn:massflow}, one thus sees
that $M$ is conserved along a streamline, such that
\begin{equation}
    \frac{dM}{dt} = 0,
    \label{eqn:LtM}
\end{equation}
which prohibits the possibility of streamline crossing.

The continuity equation has the form
\begin{equation}
    \frac{d\rho}{dt} = - \left(\frac{2g_2}{r} + H  \right) \rho,
    \label{eqn:continuity}
\end{equation}
where we have defined the velocity gradient
\begin{equation}
    H(r,t) \equiv \frac{\partial g_2}{\partial r}.
    \label{eqn:vel_grad}
\end{equation}
Euler's equation is given by
\begin{equation}
    \frac{dg_2}{dt} = - \frac{M}{r^2} + \tfrac{1}{3}\Lambda r,
    \label{eqn:euler}
\end{equation}
the integral of which along a streamline gives the Bernoulli equation,
\begin{equation}
    \tfrac{1}{2}g_2^2 -\left( \frac{M}{r} + \tfrac{1}{6}\Lambda r^2 \right) = \tfrac{1}{2}\left( g_1^2 - 1 \right).
    \label{eqn:bernoulli}
\end{equation}
This provides the physical interpretation of $g_1$ discussed
above. Indeed, by applying the comoving derivative
\eqref{eqn:comoving_deriv} to \eqref{eqn:bernoulli}, and using
\eqref{eqn:euler}, one finds the useful result,
\begin{equation}
 \frac{dg_1}{dt}=0,
    \label{eqn:Ltg1}
\end{equation}
which demonstrates that $g_1$ is conserved along a streamline.

It is most natural to specify the initial data for the above set of
equations in terms of the density and 3-velocity profiles, $\rho(r,t_i)$
and $g_2(r,t_i)$, at some initial time $t_i$. From these one can
calculate $M(r,t_i)$ and $g_1(r,t_i)$, which are then conserved along
the streamlines.



\subsection{Streamline equations}

The equations \eqref{eqn:vel} and \eqref{eqn:bernoulli} can be solved
analytically using elliptic integrals to obtain the position $r$ at
some time $t$ of a fluid particle, given its position $r_i$ at some
initial time $t_i$. It is often simpler, however, instead to solve
numerically the system of first-order ordinary differential equations
\eqref{eqn:vel} and \eqref{eqn:euler}, namely \citep{Dabrowski2000}:
\begin{align}
       \frac{dr}{dt} &= g_2, \label{eqn:str_sys_u} \\
        \frac{dg_2}{dt} &= -\frac{M(r_i)}{r^2} + \frac{\Lambda}{3}r,    \label{eqn:str_sys}
\end{align}
where $r_i$ is the position of the fluid particle on some streamline
at an initial time $t_i$. These equations can be integrated
simultaneously to find the position and velocity of the given fluid
particle at a later time $t$, using the fact that $M(r_i)$ is constant
on a given streamline. The fluid density $\rho(r,t)$ and velocity
gradient $H(r,t)$ are obtained by performing the numerical
differentiation in equations \eqref{eqn:dMdr} and
\eqref{eqn:vel_grad}. Hence, given some initial conditions for the
density and velocity distributions, $\rho(r, t_i)$ and $g_2(r, t_i)$,
the entire system is determined.

\subsection{Initial conditions}
\label{sec:ic}

We demand that the void has grown from primordial fluctuations in the
early universe. At such early epochs, it is valid to linearise the field
equations around a homogeneous cosmology, which yields two solutions:
a growing mode and a decaying mode. By demanding that the decaying
mode is absent, and assuming a flat-$\Lambda$ background cosmology,
one finds that the initial velocity and mass distributions are related
by \citep{Dabrowski2000}
\begin{equation}
    g_2(t_i, r) = \frac{2r}{3H_i} \left( 2H_i^2 - \frac{M_i}{r^3} - \frac{\Lambda}{6} \right).
    \label{eqn:ic_flat_L}
\end{equation}
This is equivalent to imposing the following relationship between
the initial velocity and density distributions
\begin{equation}
    \rho (t_i, r)=\frac{3H_i}{8\pi} \left( 4H_i - \frac{2g_{2,i}}{r} - H - \frac{\Lambda}{3H_i} \right).
    \label{eqn:ic_rho_flat_L}
\end{equation}

\subsection{Photon equations}
\label{sec:photon_eq}

The photon trajectory can be parameterised using $t$, such that it is
defined by $r(t)$ and $\phi(t)$, where $\phi(t)$ is the azimuthal angle in spherical coordinates. Without loss of generality, we may
assume that the trajectory lies in the $\theta=\pi/2$ plane. The
requirement that the trajectory is null leads to the conditions
\begin{equation}
  \begin{aligned}[b]
      \frac{dr}{dt} &= g_1 \cos \chi + g_2,   \\
     \frac{d\phi}{dt} &= \frac{\sin \chi}{r},
    \end{aligned}
    \label{eqn:gen_phot_path}
\end{equation}
where $\chi$ is the angle, as measured by observers comoving with the fluid,
between the photon path and centre of
the void (which we set to lie at the origin $r=0$). The
geodesic equations then determine how $\chi$ evolves with time:
\begin{equation}
    \frac{d\chi}{dt} = \sin \chi \left[ \cos \chi \left( H -
      \frac{g_2}{r} \right) - \frac{g_1}{r} \right].
    \label{eqn:dchidt}
\end{equation}
These equations are sufficient to calculate the position of the photon
along its trajectory, and are easier to integrate numerically than the
usual second-order geodesic equations.  The trajectory is determined
by an initial set of data $r_i$, $\phi_i$ and $\chi_i$.  For most
calculations, however, the data are provided in the form of the
observer's position and an angle on the sky $\chi$, and the equations
are then run backwards in time to take the photon back through the
void.

The remaining content of the geodesic equations determines the
evolution of the photon frequency $\omega$, as measured by a comoving
observer:
\begin{equation}
    \frac{d\omega}{dt} = -\omega \left(H \cos^2\chi +\frac{g_2}{r} \sin^2\chi  \right).
    \label{eqn:dwdt}
\end{equation}
One can show from the above equations that the angular momentum of the
photon, $L= -r^2 \omega\,d\phi/dt = -\omega r \sin \chi$, is
conserved.

To calculate the effect of the void on a CMB photon, we first write
\begin{equation}
    g_2(t,r)= rH_e(t) + \Delta(t,r),
    \label{eqn:D_defn}
\end{equation}
where $H_e(t)$ is the Hubble function in the exterior Universe at time
$t$, and $\Delta$ is thus the difference between the equivalent fluid
velocity in the unperturbed Universe and in the void. One may then
show that the physically measurable CMB temperature decrement due to
the void is given by \citep{Lasenby1999}
\begin{equation}
    \frac{\Delta T}{T}  = e^{-\epsilon}-1 \approx -\epsilon,
    \label{eqn:DT}
\end{equation}
where the small quantity $\epsilon$ is defined by
\begin{equation}
    \epsilon = \int_{t_1}^{t_2} dt \left( \frac{\partial \Delta}{\partial r} \cos^2 \chi + \frac{\Delta}{r} \sin^2 \chi  \right).
    \label{eqn:epsilon}
\end{equation}
This integral is evaluated along the photon path between the
time the photon enters the void $(t_1)$ and the time it leaves
$(t_2)$. The function $\epsilon$ is small, since the contribution to
the integral from near the void centre tends to cancel the
contributions from further out.  The main effect producing a non-zero
$\epsilon$ is essentially the evolution of $\Delta$ with time.

\subsection{Void model}
\label{sec:our_model}

Our void model is based on that used to model a cluster in
\cite{Dabrowski1999}. The nature of the void is determined by
specifying the velocity distribution $g_2(r,t_i)$ at some initial time
$t_i$ in terms of four parameters $H_i$, $R_i$, $a$ and $m$. Here,
$H_i \equiv H_e(t_i)$ is the background Hubble parameter at $t=t_i$, $R_i$ is the
initial size of the perturbed region, and $a$ is the velocity gradient
at the origin (which determines the magnitude of the
perturbation). For $r < R_i$, the fluid velocity is described by a
polynomial in $r$ of order $2m + 1$, and the first $m$ radial
derivatives are matched at the boundaries, $r=0$ and $r=R_i$.  For $r
> R_i$, the fluid velocity is that of the background
$g_2(r,t_i)=rH_i$. The initial density profile $\rho(r,t_i)$ is
then determined using equation \eqref{eqn:ic_rho_flat_L}, such that
the density is a polynomial of order $2m$. One can show that the
resulting initial density profile is compensated, and hence remains
compensated for all time. Consequently, in the external region
($r>R_i$) the fluid evolves as a homogeneous FRW universe.
Placing observers in this region allows for unambiguous calculations
of the CMB perturbation caused by the void. Once the initial velocity
and density profiles are defined, the evolution of the fluid is then
completely determined.

A velocity gradient at the origin that is slightly greater than that
of the unperturbed universe, so that $a > H_i$, leads to the
formation of a void (and, conversely, setting $a < H_i$ would lead to the
formation of a cluster). Also, the boundary conditions imply that when
the value of $m$ is greater than unity, the density gradient is zero
at the origin, whereas this condition is not necessarily satisfied for
$m=1$; hence we choose $m\geq 2$ for a sensible density profile.


\section{LTB model}
\label{sec:LTB}


Having discussed our own methodology for modelling voids, we now turn to an approach based on the LTB model, and establish the relations between quantities in the two models.

The LTB metric \citep{Lemaitre1933, Tolman1934, Bondi1947}
describes a spherically-symmetric pressureless system, and may be
written in the form
\begin{equation}
    ds^2 = d\hat{t}^2 - \frac{(\partial_{\hat{r}} R)^2}{1+2E(\hat{r})} - R^2 d\Omega^2,
    \label{eqn:LTB_metric}
\end{equation}
where $\hat{r}$ is a comoving radial coordinate and the time
coordinate $\hat{t}$ coincides with the proper time measured by
observers comoving with the fluid. The function $R$ depends, in
general, on both $\hat{t}$ and $\hat{r}$, and the function
$E(\hat{r})$ determines the so-called `curvature profile' of the
system, and may be specified arbitrarily, provided $E(\hat{r}) >
-\frac{1}{2}$.

As discussed in \cite{Kim2016a}, one may transform the line-element
\eqref{eq:generalmetric} used in our tetrad-based approach into
the LTB line-element \eqref{eqn:LTB_metric} via the coordinate transformation
\begin{equation}
  t=\hat{t}, \quad r=R(\hat{r},\hat{t}), \quad \text{where} \quad
  \frac{\partial R}{\partial \hat{t}}=g_2,
  \label{eqn:transf}
\end{equation}
and in so doing one makes the further identification $g_1^2 = 1+
2E(\hat{r})$, which confirms the usual alternative interpretation of
$E(\hat{r})$ as the energy per unit mass of a fluid particle (after
subtracting its rest mass). One may also show that partial derivatives
in the two coordinate systems are related by
\begin{align}
\frac{\partial}{\partial \hat{t}}
        &= \frac{\partial}{\partial t} + g_2 \frac{\partial}{\partial r}
        = \frac{d}{dt},
    \label{eqn:d_dt_comov} \\
\frac{\partial}{\partial \hat{r}} &= \frac{\partial R}{\partial \hat{r}} \frac{\partial}{\partial r}.\label{eqn:d_dr_comov}
\end{align}
Since $\frac{d}{dt}$ is the convective derivative, the transformation
\eqref{eqn:transf} is naturally interpreted as moving from a Eulerian
to a Lagrangian description of the fluid motion.

In LTB coordinates, one finds $\partial_{\hat{t}} M=0$, so that
$M=M(\hat{r})$, which is interpreted as the mass contained within the
comoving radius $\hat{r}$. The remaining Einstein equations become
\begin{align}
    \partial_{\hat{r}}M(\hat{r}) &=4 \pi R^2 \rho \,\partial_{\hat{r}}R, \label{eqn:LrM_comov}\\
    \left( \partial_{\hat{t}}R \right)^2 & = 2E(\hat{r}) + \frac{2 M(\hat{r})}{R} + \tfrac{1}{3}\Lambda R^2.
    \label{eqn:vel_LTB}
\end{align}
The latter is the LTB equivalent of the Bernoulli equation
\eqref{eqn:bernoulli}, for which the solution $R(\hat{r}, \hat{t})$
can be written in integral form as
\begin{equation}
    \hat{t}-\hat{t}_B(\hat{r}) = \int_0^{R(\hat{r}, \hat{t})}
    \frac{dA}{\sqrt{2E(\hat{r}) + \frac{2 M(\hat{r})}{A} +
        \tfrac{1}{3}\Lambda A^2}},
    \label{eqn:t_LTB}
\end{equation}
where the function $\hat{t}_B(\hat{r})$ is known as the bang-time,
which is interpreted as a Big-Bang singularity surface at which $R$
vanishes, i.e. $R(\hat{r},\hat{t}_B(\hat{r}))=0$, and may be chosen arbitarily.

In the LTB model, one is thus required to specify three arbitrary
time-independent functions. The functions $E(\hat{r})$ and
$\hat{t}_B(\hat{r})$ are usually interpreted as determining the nature
of the system, and the remaining gauge freedom in redefining the
radial coordinate is usually removed by specifying $M(\hat{r})$.  In
this case, \eqref{eqn:t_LTB} can then be solved for $R(\hat{r},
\hat{t})$, so that metric \eqref{eqn:LTB_metric} is fully determined,
and the corresponding density $\rho(\hat{r},\hat{t})$ is found from
\eqref{eqn:LrM_comov}. A common alternative gauge-fixing procedure is
instead to specify $R(\hat{r},\hat{t}_0)$, where $\hat{t}_0$ denotes
the current epoch $\hat{t}_0$. In this case, \eqref{eqn:t_LTB} is then
solved at $\hat{t}_0$ for $M(\hat{r})$, before proceeding as before.
It is worth noting that, in order for no decaying mode to be present,
one requires $\hat{t}_B(\hat{r})$ not to be spatially varying, and one
may choose $\hat{t}_B=0$ without loss of generality \citep{Zibin2008}.


\section{Modelling the Draco void}

We now wish to compare our tetrad-based approach with the LTB model in in the context of modelling a specific void structure. Our main aim will be to show that in comparing void models with CMB data, both the underlying density and velocity profiles of the void are important, and should both be considered. We also wish to demonstrate, however, that the tetrad-based approach is simpler and more intuitive in some respects than the LTB model, although the latter does of course remain a valid approach. Indeed, when the issues related to density versus velocity profile are set aside, the quantitative differences that arise between the two methods in modelling the cosmic void considered below are only at the few per cent level. 

The particular approach we consider is that adopted by \cite{Finelli2016} (hereinafter FGKPS). They used the LTB metric to model
compensated voids and showed that a large void can lead to a
significant temperature decrement in the CMB. Indeed, in the first version of their paper \citep{Finelli2014}, and following their
identification of a projected underdensity in the WISE-2MASS all-sky
infrared galaxy catalogue aligned with the CMB Cold Spot direction,
they originally stated that a supervoid of size $\sim 400$~Mpc and
depth (fractional overdensity) $\delta \sim -0.10$ can explain the
Cold Spot temperature decrement of $\Delta T \sim 150 \mu$K; they also
stated that this decrement is mostly due to the Rees--Sciama (RS)
effect \citep{Finelli2014}, rather than the linear integrated Sachs-Wolfe (ISW) effect. Nonetheless, it
was later shown by \cite{Zibin2014} and \cite{Nadathur2014}, again
using the LTB approach, that the ISW effect dominates and that a void
of these proportions is not capable of producing such a large CMB
decrement. In the later version of their paper, FGKPS accept that the
Cold Spot decrement is difficult to explain by the presence of a
single supervoid only, but drew attention to another sky area where a large underdensity in the
projected WISE-2MASS galaxy map (which they call the Draco supervoid)
can account for the CMB decrement observed in that direction.

In this section, we therefore summarise the LTB approach used by FGKPS to describe the Draco supervoid and then in the following section compare the results of this model with what is obtained using our own tetrad-based model.

\subsection{Modelling the Draco void using an LTB approach}

To model the Draco void, FGKPS choose the curvature profile to have
the form
\begin{equation}
    E(\hat{r})= E_0 \hat{r}^2 \exp \left(-\frac{\hat{r}^2}{\hat{r}_0^2} \right).
    \label{eqn:E_LTB}
\end{equation}
FGKPS do not specify their choice for the bang-time
$\hat{t}_B(\hat{r})$, but we presume that it is set to a
constant (which one may take to be zero) so that their model contains
no decaying mode. FGKPS also do not specify their gauge
choice, but reference is made to an earlier work \citep{Garcia-bellido2008} in which
they fix the gauge by setting $R(\hat{r},\hat{t}_0)=\hat{r}$. From
\eqref{eqn:LrM_comov}, this is equivalent to setting $M(\hat{r}) =
\frac{4\pi}{3}\hat{r}^3\rho(\hat{r},\hat{t}_0)$, where $M(\hat{r})$ is
determined from \eqref{eqn:t_LTB}.

\begin{figure}
\centering
    \includegraphics[width=\columnwidth]{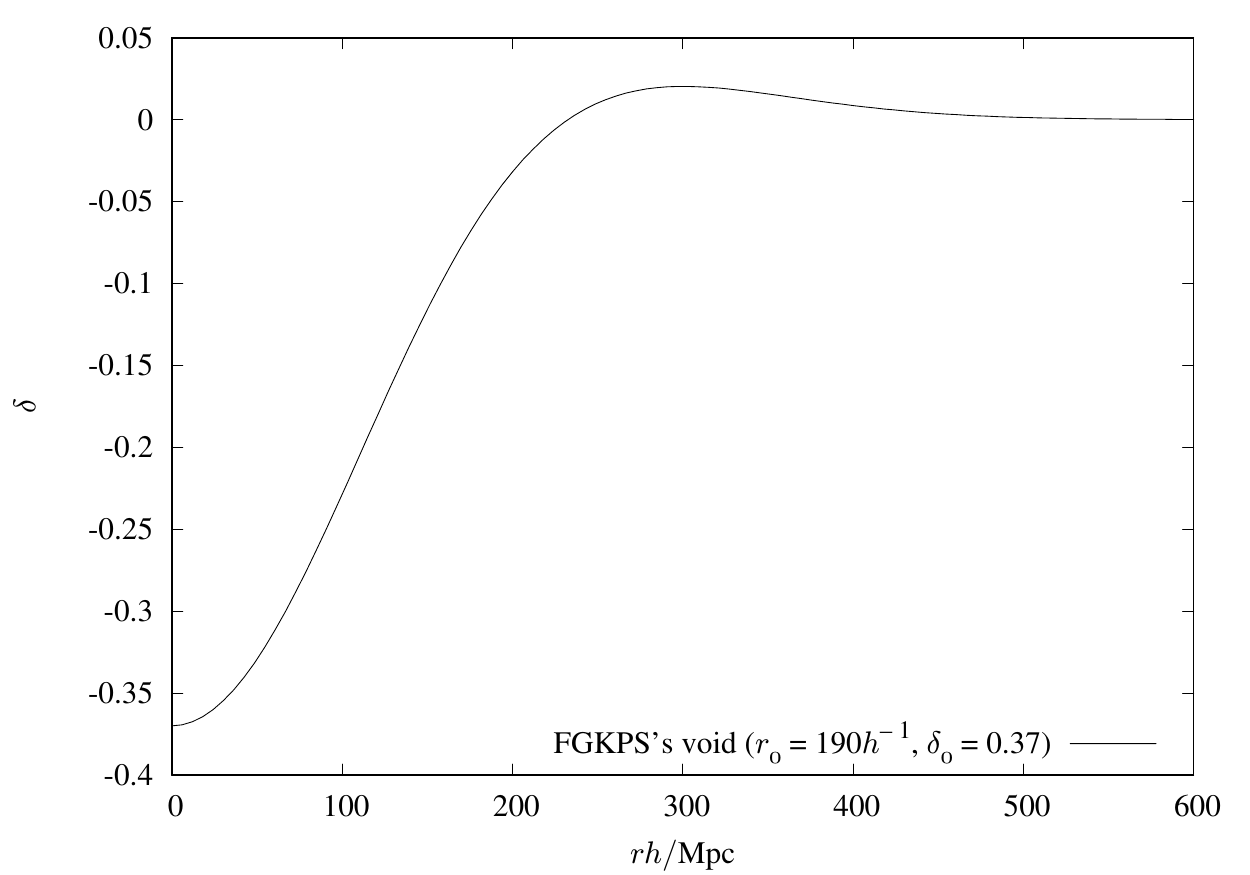}
    \caption{Present-day density contrast used by \protect\cite{Finelli2016} to model the Draco supervoid.}
    \label{fig:szapudi_delta}
\end{figure}

FGKPS also appear to
fix the present-day density contrast, which as pointed out by \cite{Nadathur2014} could over-constrain their
model, and we discuss this aspect further below. In particular, they assume the curvature profile
\eqref{eqn:E_LTB} to correspond to a perturbation $\Phi$ in the
synchronous gauge in a spatially-flat $\Lambda$CDM model. Treating the
perturbation as linear gives rise to the present-day metric
perturbation $\Phi(\hat{r}) = \Phi_0 \exp (-\hat{r}^2/\hat{r}_0^2)$.
Taking the growing mode and using the Poisson equation, FGKPS obtain
the present-day density contrast
\begin{equation}
    \delta(\hat{r}) = - \delta_0 \left( 1-\frac{2\hat{r}^2}{3 \hat{r}_0^2} \right) \exp \left(-\frac{\hat{r}^2}{\hat{r}_0^2} \right),
    \label{eqn:delta_sz}
\end{equation}
which is plotted in Fig.~\ref{fig:szapudi_delta} and ensures that the
void is compensated at infinity.

This density contrast (projected onto the sky) and the CMB decrement
produced by the RS and ISW effects calculated from the metric
perturbation $\Phi$ are then used in a simultaneous $\chi^2$ fit to
the void in the WISE-2MASS galaxy catalogue and Planck CMB data to
estimate the three parameters, $\delta_0$, $r_0$ and $z_0$, where
$z_0$ is defined as the redshift at the centre of the void. The
best-fit values and 68 per cent confidence limits were found to be
\begin{equation}
  \begin{aligned}[b]
    \delta_0 &= 0.37^{+0.22}_{-0.12} \, , \\
    r_0 &= 190^{+39}_{-27} \; \mathrm{Mpc}/h, \\
    z_0 &= 0.15 ^{+0.04}_{-0.05},
    \end{aligned}
    \label{eqn:para_for_draco}
\end{equation}
where $h$ is defined such that the current Hubble parameter is given
by $H_0 =100h \; \mathrm{km\,s^{-1}Mpc^{-1}} $.

\subsection{Presence of decaying mode}
\label{sec:decaying_mode}

If FGKPS do indeed specify both the gauge condition
$R(\hat{r},\hat{t}_0)=\hat{r}$ and the present-day density contrast
\eqref{eqn:delta_sz}, then it will not be possible, in general, to
satisfy the constraint \eqref{eqn:t_LTB} with $\hat{t}_B(\hat{r})$
being equal to a constant (usually zero). Consequently, the void model
would contain some contribution from a decaying mode, and thus be
incompatible with the standard picture of cosmological structure
formation \citep{Zibin2008, Nadathur2014}.

To demonstrate this possibility, we calculate the quantity
\begin{equation}
    \hat{t}'(\hat{r})
= \int_0^{\hat{r}} \frac{dA}{\sqrt{2E(\hat{r}) + \frac{2 M(\hat{r})}{A} + \tfrac{1}{3}\Lambda A^2}},
    \label{eqn:tdashed_LTB}
\end{equation}
in which $E(\hat{r})$ is given by \eqref{eqn:E_LTB} and
\begin{equation}
    M(\hat{r}) = \frac{4 \pi \hat{r}^3}{3} \bar{\rho}(\hat{t}_0)
    \left( 1+\delta(\hat{r}) \right),
    \label{eqn:M_LTB}
\end{equation}
where $\bar{\rho}(\hat{t}_0)$ is the present-day background FRW
density and the present-day density contrast $\delta(\hat{r})$ is given by
\eqref{eqn:delta_sz}. If this specification were consistent with the
absence of a decaying mode contribution, then $\hat{t}'(\hat{r}) $
should be constant and equal to $\hat{t}_0$, which is
13.5~Gyr for the assumed background cosmology.

In addition, FGKPS do not specify the value of $E_0$ used to set the amplitude of their curvature profile in \eqref{eqn:E_LTB}. We therefore try two different methods to determine the value of $E_0$. The first is to find the value of $E_0$ which minimises the rms variation of $\hat{t}'(\hat{r})$ over the range $0<\hat{r}<600$ Mpc.  Using the best-fit values of $\delta_0$ and $\hat{r}_0$ for the Draco supervoid given in \eqref{eqn:para_for_draco}, we find that the rms variation is minimised when $E_0 \hat{r}_0^2 = 5.3 \times 10^{-4}$.  The second method is to treat the perturbation as having linearly grown (as outlined in \cite{Zibin2014}). This results in $E_0 \hat{r}_0^2 = 4.78 \times 10^{-4}$. One would expect the nonlinear growth to deviate from this however, as the density contrast is fairly large.

\begin{figure}
    \centering
    \includegraphics[width=\columnwidth]{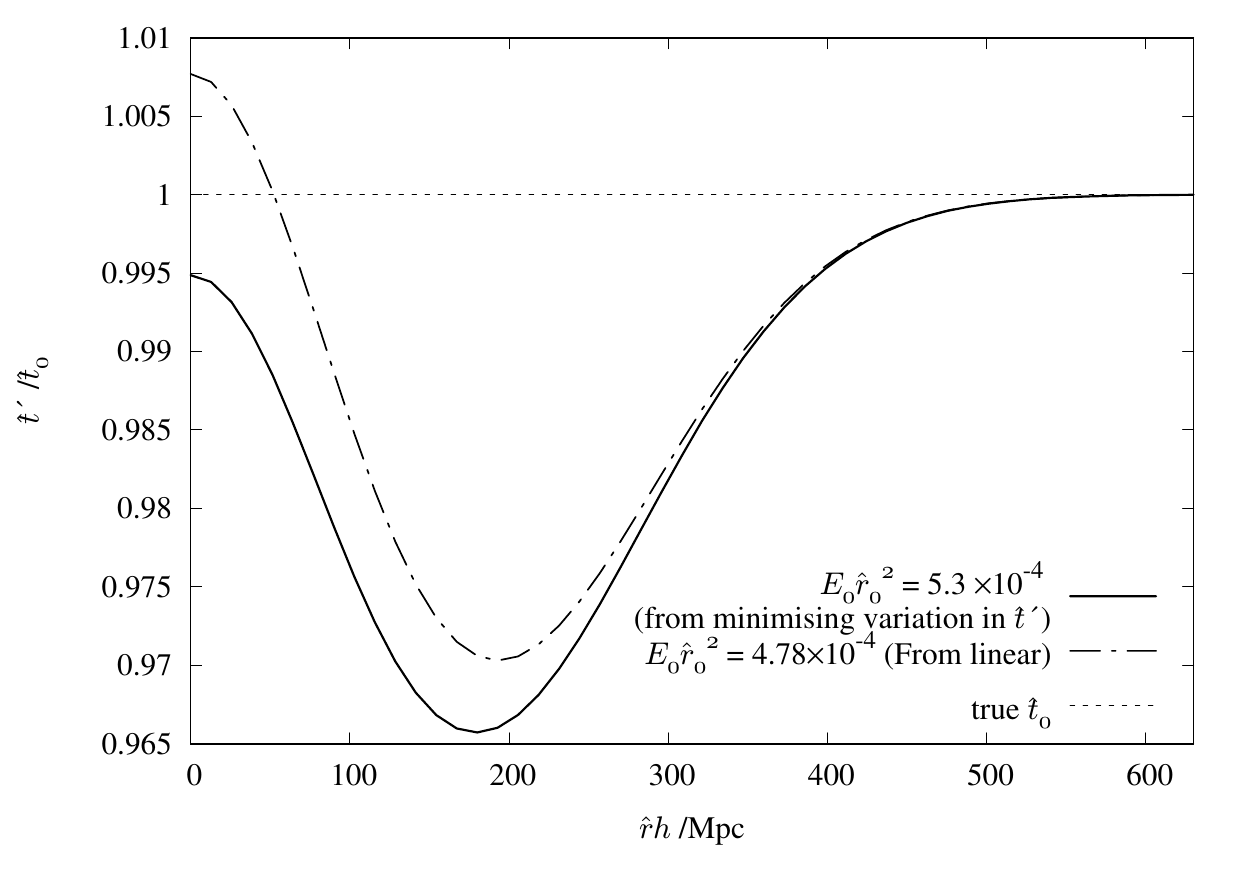}
    \caption{Plot of $\hat{t}'$ using different values of $E_0$, using parameters for the Draco supervoid \eqref{eqn:para_for_draco}. The variation is at its smallest when $E_0\hat{r}_0^2 = 5.3 \times 10^{-4}$. As $\hat{r} \rightarrow \infty$, $\hat{t}' \rightarrow \hat{t}_0$ for all values of $E_0$, as expected.}
    \label{fig:t_var}
\end{figure}

The corresponding ratio $\hat{t}'(\hat{r})/\hat{t}_0$ is plotted in Fig.~\ref{fig:t_var} for each value of $E_0$, and clearly neither is constant and equal to unity.
Indeed, $\hat{t}'(\hat{r})$ exhibits a $\sim 2$ per cent variation
about a mean value of $\approx 13.2$~Gyr. We may thus conclude that
the resulting LTB model must have some contribution from a decaying
mode, although the difference between the imposed present-day density
profile and one that contains no decaying mode contribution is only at
the level of a few per cent. This is in agreement with level of
inconsistency found by \cite{Nadathur2014}, who instead estimated the
value of $E_0$ in the FGKPS model by minimising the difference
$\hat{t}'(\hat{r})-\hat{t_0}$ at the single point $\hat{r}=0$
(Nadathur, private communication).  Our findings are also consistent
with those of \cite{Zibin2014}, who finds that the present-day density
profile imposed by FGKPS differs from that obtained from
\eqref{eqn:LrM_comov} in a self-consistent LTB model, although again
the discrepancy is small. In the rest of the paper we choose the value of $E_0$ found by minimising the rms variation in $\hat{t}'(\hat{r})$, namely $E_0 \hat{r}_0^2 = 5.3 \times 10^{-4}$.




\section{Reproducing the Draco void}
\label{sec:mimic1}

The issues outlined in the previous section arise, in part, from the
subtleties associated with gauge-fixing in the LTB model, for which
the physical interpretation in unintuitive. In this Section, we
therefore instead model the Draco void using the tetrad-based approach
described in Section~\ref{sec:tetrad}, for which we believe the physical
interpretation is clearer. Rather than simply fitting our model
directly to the WISE-2MASS galaxy survey data and Planck CMB
observations, however, we wish to focus on the different nature of the
void characteristics in the LTB and tetrad-based
approaches. Consequently, we will compare our model instead with the
LTB model derived by FGKPS for the Draco void. In particular, we wish
to determine the CMB temperature decrement produced by a void with
present-day characteristics similar to those of the FGKPS void model
for Draco, but modelled using our approach.

As mentioned in Section~\ref{sec:our_model}, in our approach the nature of the
present-day void is determined by specifying the velocity distribution
$g_2(r,t_i)$ at some initial time in terms of the four parameters
$H_i$, $R_i$, $a$ and $m$. We choose here to set these conditions
at $z=10^3$, since the perturbations are safely within the linear
regime at this epoch. The initial density distribution is then
determined by the requirement that the decaying mode is absent in the
linearised limit, as discussed in Section~\ref{sec:ic}.  Following FGKPS, we
assume a spatially-flat $\Lambda$CDM background cosmology with
$\Omega_{\Lambda} = 0.7$, for which $H_i =1.735 \times 10^{6}h
\,\mathrm{km\,s^{-1}Mpc^{-1}}$. To determine the appropriate values
for the remaining parameters $R_i$, $a$ and $m$, one has the choice of
attempting to reproduce the present-day distribution in the FGKPS
Draco void model of either: density, velocity, or both. We consider
each of these in turn. As we will see, these options lead to very
different CMB temperature decrements.

\subsection{Reproducing the density profile}
\label{sec:fitrho}

We begin by choosing the initial velocity perturbation parameters
$R_i$, $a$ and $m$ such that the resulting present-day density profile
of the void $\rho(r,t_0)$ is as similar as possible to that of the FGKPS Draco void
model, which is plotted in Fig.~\ref{fig:szapudi_delta}.

\begin{figure}
\centering
 \includegraphics[width=\columnwidth]{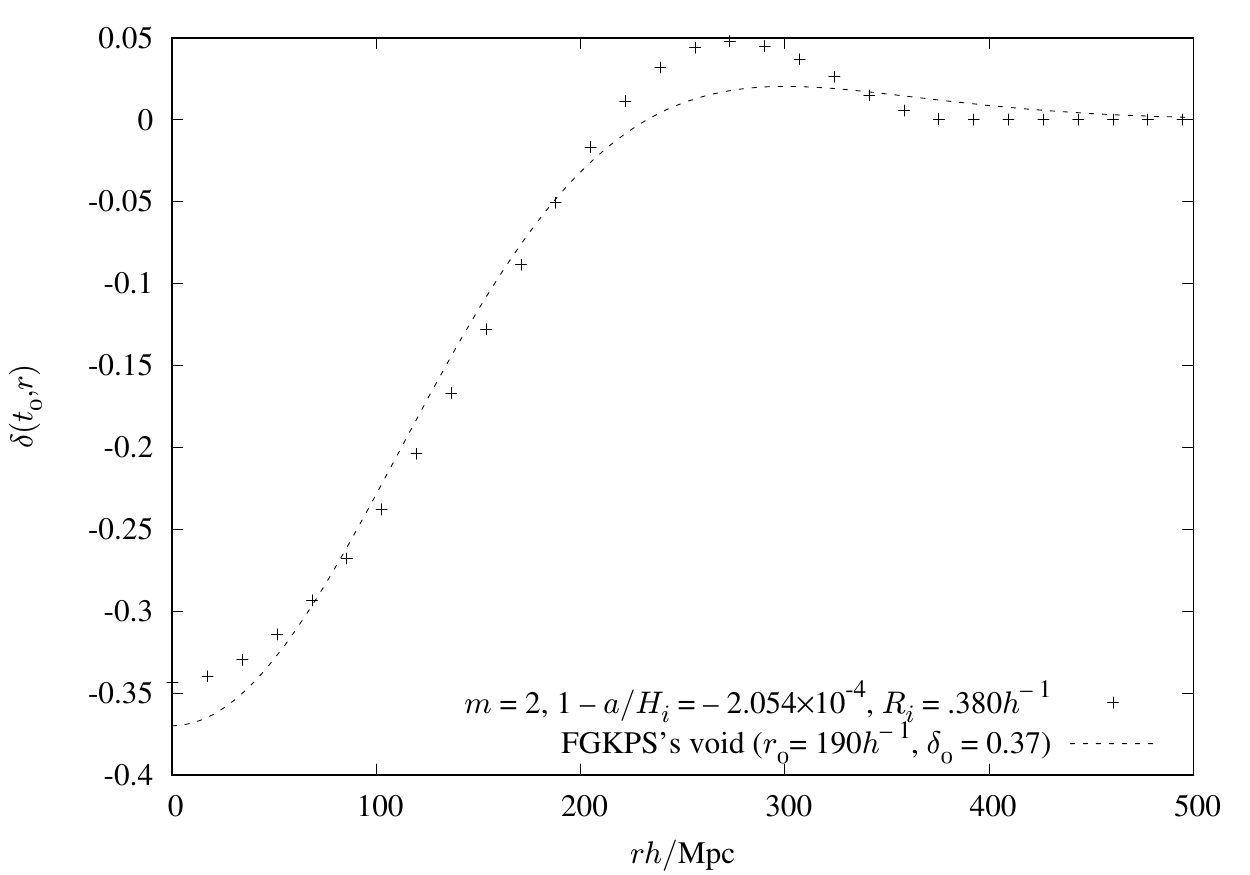}
 \caption{Void density profile at $t=t_0$ in our model (crosses),
   resulting from an initial velocity perturbation at $z=10^3$ with
   parameters chosen to reproduce the FGKPS void density profile
   (solid line) as closely as possible.
    \label{fig:rho_compare}}
\end{figure}

This is achieved by performing a numerical optimisation in which the
values of $R_i$, $a$ and $m$ are varied. For each set of parameter
values, the resulting present-day density profile is evaluated at 40
equally-spaced points in the range $0 < r < 500$~Mpc and compared with
the corresponding values in the FGKPS Draco void density profile.  The
optimal values of the parameters are those that minimise and sum of
the squares of the differences in the two profiles. The resulting
optimal present-day density profile is compared with the FGKPS density
profile in Fig.~\ref{fig:rho_compare}. Given the very different
underlying physical models, there is reasonably close agreement
between the density profiles. In particular, we find that lower values
of the parameter $m$ are preferred, since they result in a smoother
density profile in our model; we thus use $m=2$ for the remainder of
this section.

Since our primary interest, however, is in the CMB temperature
decrement produced by our void, it is worth recalling from
Section~\ref{sec:photon_eq} that the main contribution to this effect is the
evolution of the difference $\Delta$ between the equivalent fluid
velocity in the external universe and in the void, as the CMB photon
traverses it. Consequently, it is of interest also to compare the
velocity distribution of our void to that of the FGKPS void.

\begin{figure}
\centering
    \includegraphics[width=\columnwidth]{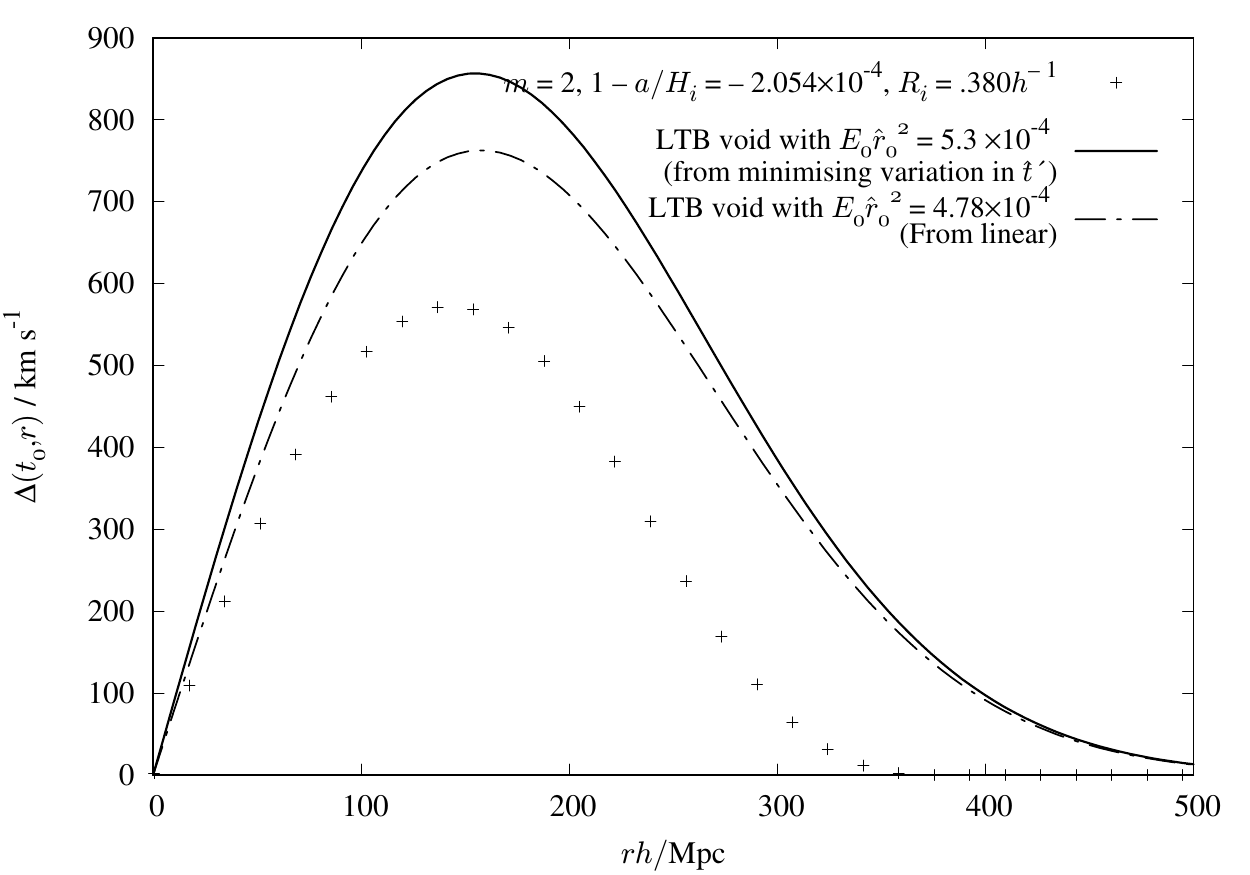}
    \caption{Difference $\Delta \equiv g_2-rH_e$ at $t=t_0$ between
      the fluid velocity in the void and in the external universe for
      our void model (crosses) and the LTB void model of FGKPS for two
      different values of $E_0$ (solid and dotted lines).
    \label{fig:v_compare_m2}}
\end{figure}

To determine the velocity distribution of the FGKPS void, we note from
\eqref{eqn:transf} that the LHS of \eqref{eqn:vel_LTB} gives the square
of the fluid velocity in the LTB model at any given epoch. Using the
forms for $E(\hat{r})$ and $M(\hat{r})$ given in \eqref{eqn:E_LTB} and
\eqref{eqn:M_LTB}, respectively, and employing the gauge condition
$R(\hat{r},\hat{t}_0)=\hat{r}$, one thus obtains the velocity profile
at the current epoch. The corresponding $\Delta$ profile is plotted
Fig.~\ref{fig:v_compare_m2} for two different values of the parameter
$E_0$ that defines the amplitude of the LTB curvature profile (solid
and dotted lines), together with the $\Delta$ profile from our void
(crosses). The solid line corresponds to the value of $E_0$ determined
in Section~\ref{sec:decaying_mode}, whereas the dotted lines corresponds to a value of
$E_0$ that is a factor of 5 smaller.  Two points are worth noting from
this figure. First, the $\Delta$ profile of the LTB void is sensitive
to the value of $E_0$, although the maximum difference between the two
LTB $\Delta$ profiles is only at the $\sim 10$ per cent level.
Second, the velocity profile of our void differs substantially from
that of the LTB void for either value of $E_0$, even though the
density profiles of the two voids agree reasonably well (by
construction), as shown in Fig.~\ref{fig:rho_compare}.

\begin{figure}
\centering
    \includegraphics[width=\columnwidth]{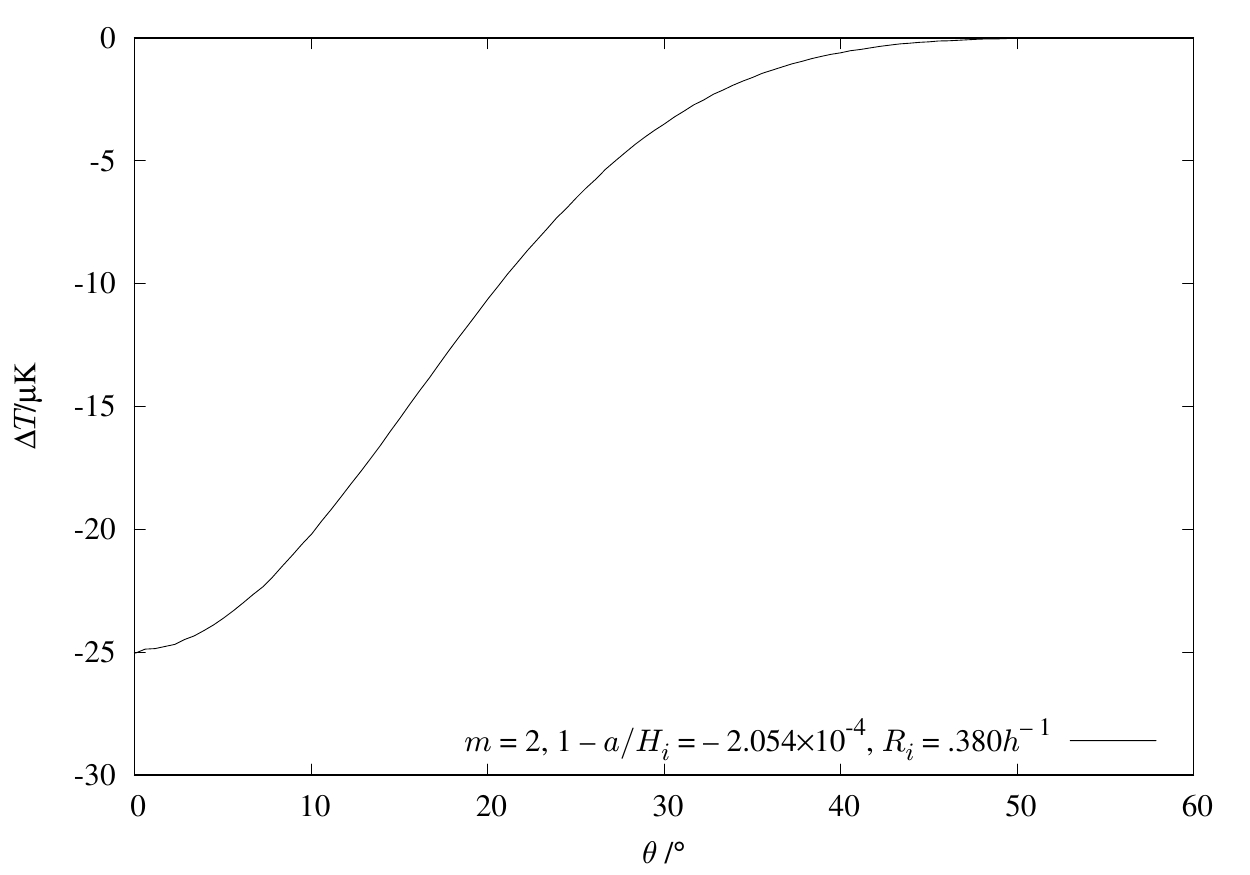}
    \caption{CMB temperature decrement for our void model that best
      reproduces the present-day density distribution of the FGKPS
      void. The background CMB temperature is taken to be 2.725K.}
    \label{fig:dT_plot_fit_void}
\end{figure}

\begin{figure}
\centering
    \includegraphics[width=\columnwidth]{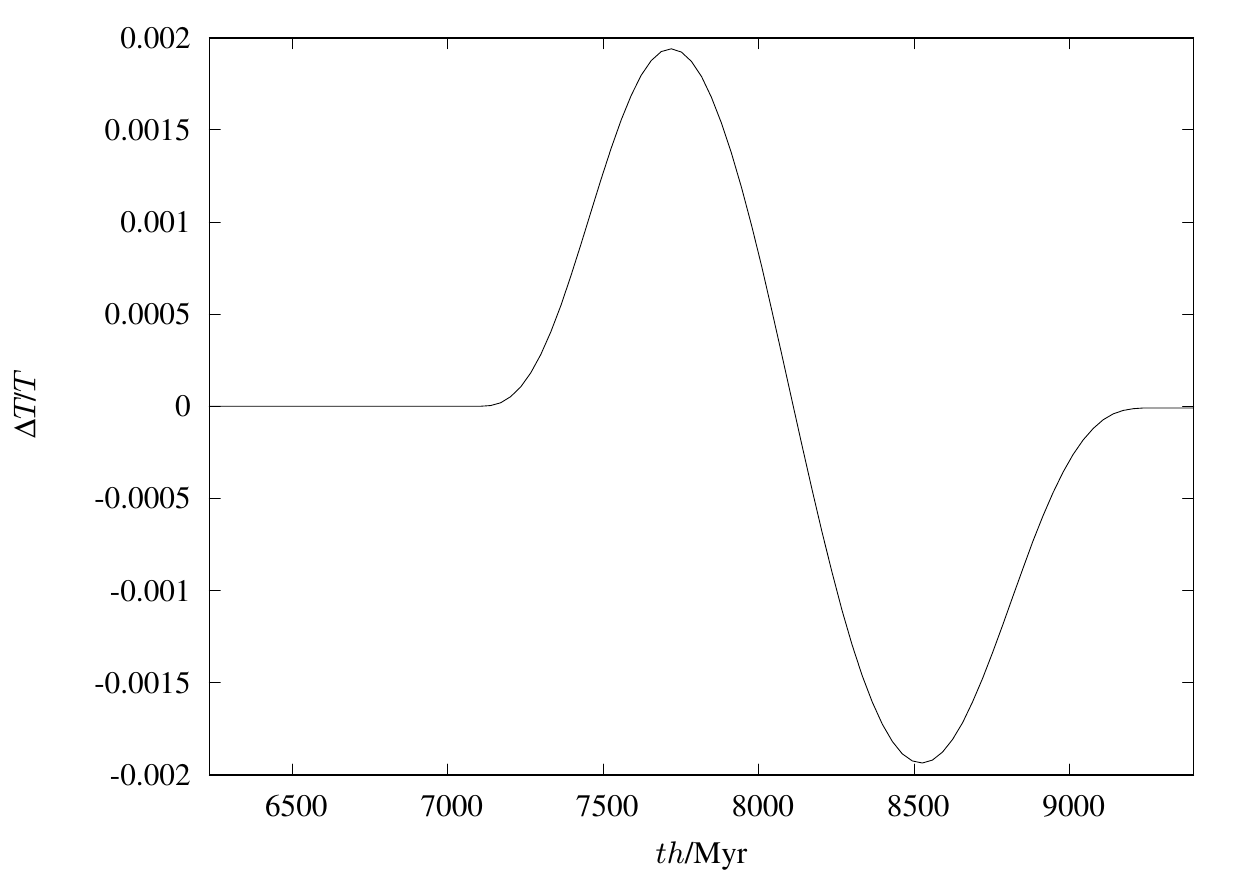}
    \includegraphics[width=\columnwidth]{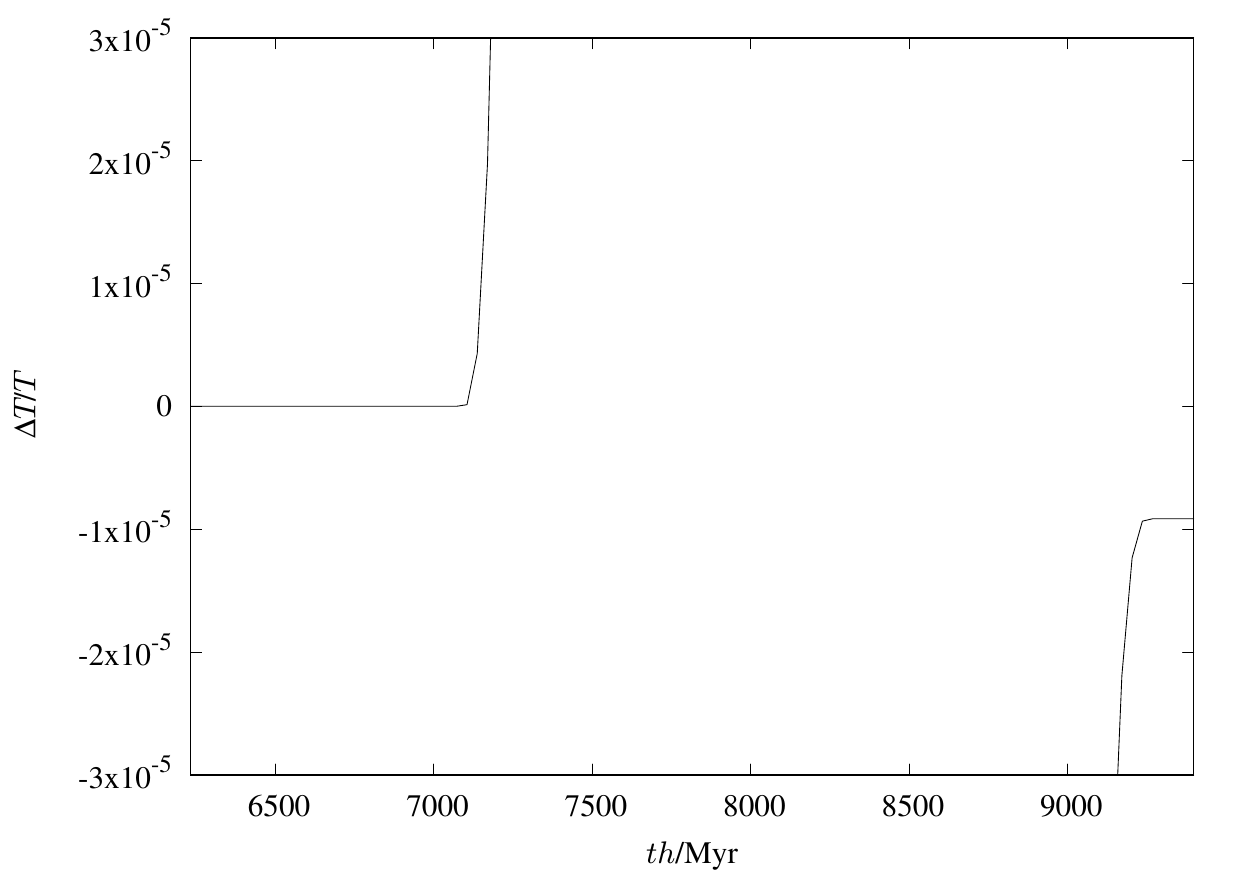}
    \caption{As in Fig.~\ref{fig:dT_plot_fit_void}, but for the
      variation of $\Delta T/T$, as measured by a comoving observer at
      each point along the path of a photon that passes through the
      centre of the void. The lower panel is identical to the upper
      panel, but is plotted on an expanded $\Delta T/T$ scale to
      illustrate the final decrement of $\Delta T/T = -9.16\times
      10^{-6}$.
    \label{fig:path_dT_void}}
\end{figure}

To determine the CMB temperature decrement produced by our void, we
consider an observer comoving with the cosmic fluid in the external
universe. The present-day distance of the observer from
the void is chosen such that the centre of the void lies at a redshift
$z_c = 0.15$. This corresponds to a comoving radial coordinate
distance $\hat{r} = 434 h^{-1}$ Mpc. Since FKGPS employ the gauge
condition $R(\hat{r},\hat{t}_0)=\hat{r}$ at $t_0$, this corresponds
simply to a non-comoving radial coordinate distance of $r=434h^{-1}$
Mpc at the current epoch. The resulting CMB temperature decrement
$\Delta T(\theta)$ is plotted in Fig.~\ref{fig:dT_plot_fit_void}, and is very similar to that obtained by FGKPS, both in terms of its angular profile and central value of $\Delta T \approx -25$~$\mu$K. In Fig.~\ref{fig:path_dT_void}, we also plot the variation of $\Delta T/T$, as measured by a comoving observer at each point along the path of a photon that passes through the centre of the void, which nicely demonstrates that the observer does indeed lie in the external universe, beyond the finite compensation radius of our void.

\subsection{Reproducing the velocity profile}
\label{sec:fitv}

Since the velocity profile of our void discussed above differs
significantly from that of the FGKPS void, an alternative approach is
instead to choose the initial velocity perturbation parameters $R_i$,
$a$ and $m$ such that the resulting present-day velocity profiles of
the two voids models are as close as possible.  This is achieved by
performing an analogous numerical optimisation to that used above.

\begin{figure}
\centering
    \includegraphics[width=\columnwidth]{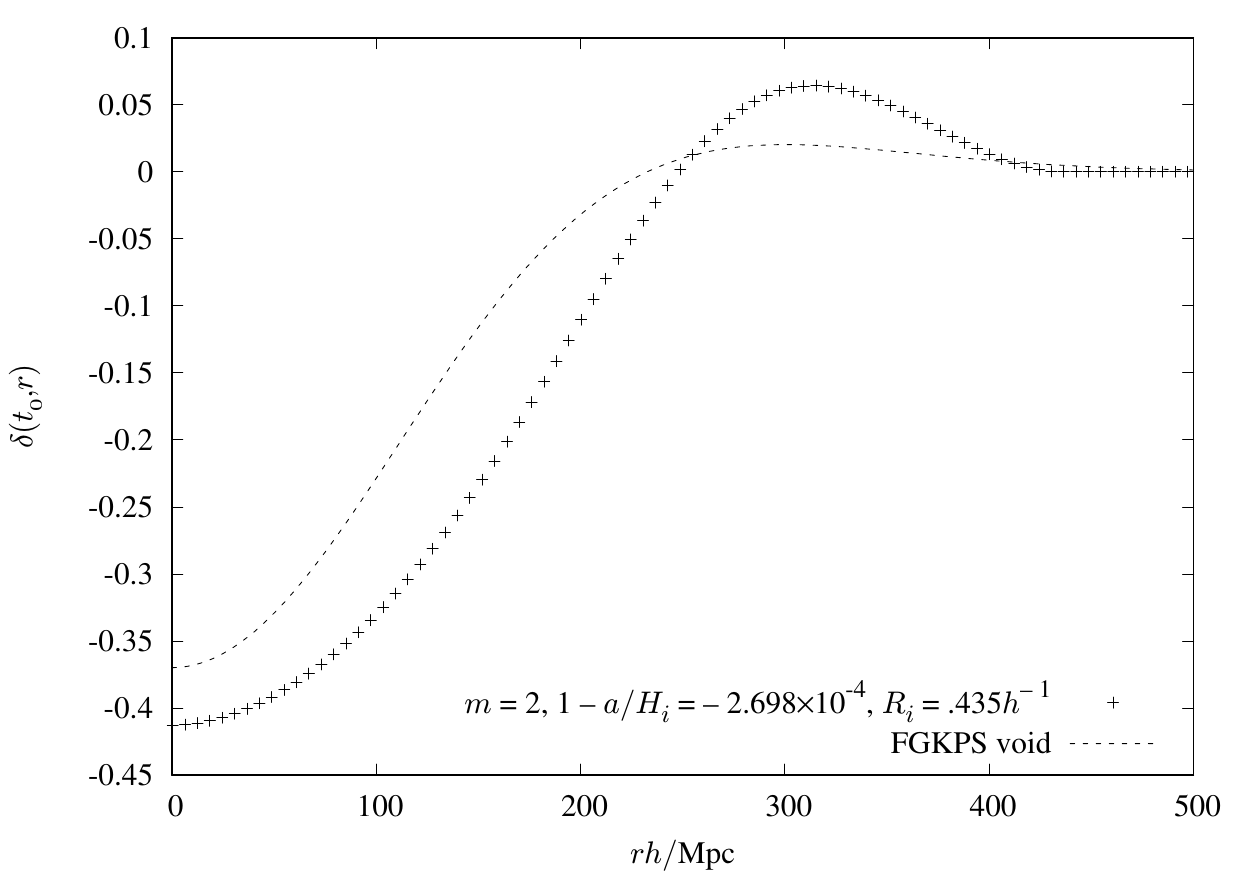} \quad
    \includegraphics[width=\columnwidth]{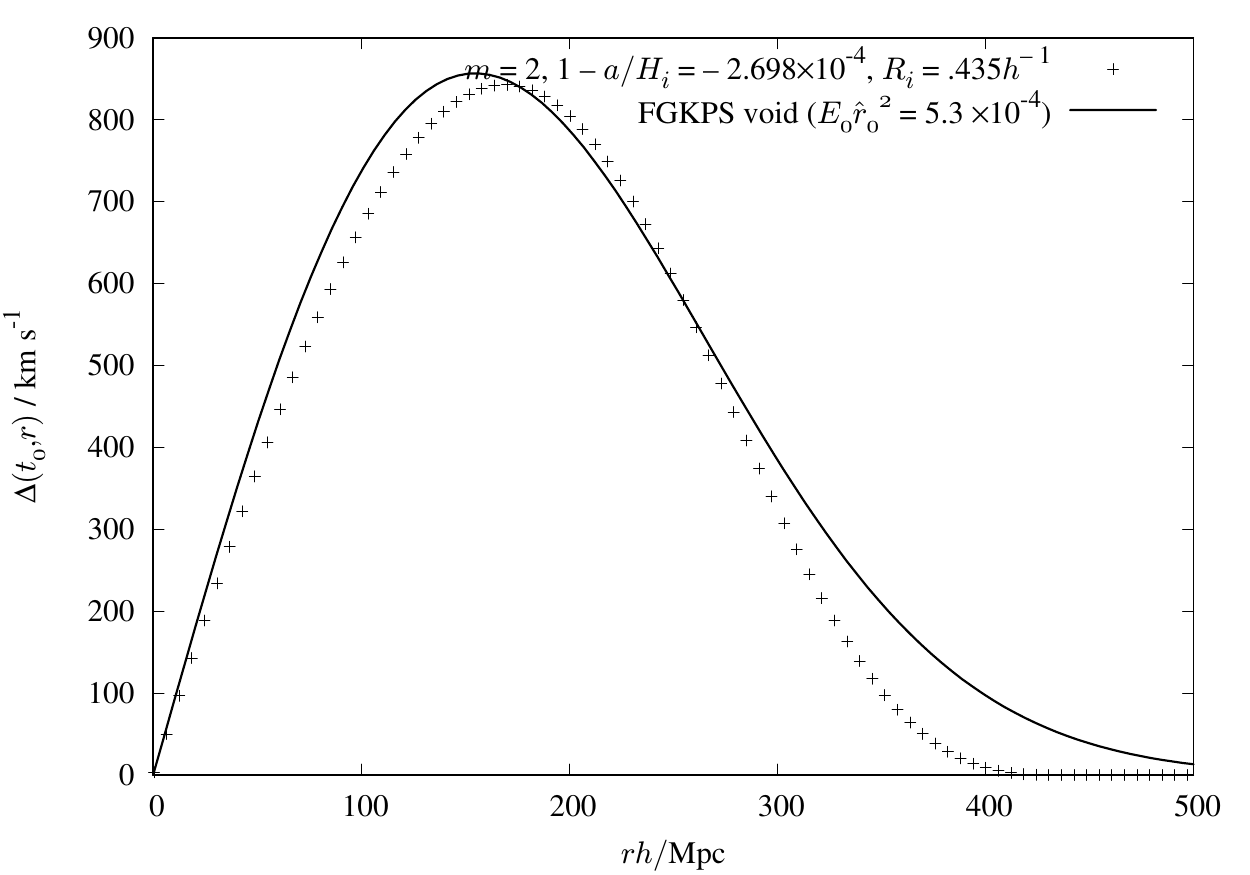}
 \caption{Void density profile (top) and velocity difference $\Delta$
   profile (bottom) at $t=t_0$ in our model (crosses), resulting from
   an initial velocity perturbation at $z=10^3$ with parameters chosen
   to reproduce the FGKPS void velocity profile (solid line, bottom
   panel) as closely as possible.}
    \label{fig:rho_v_fit_v}
\end{figure}

The resulting present-day density and velocity difference $\Delta$
profiles are plotted in Fig.~\ref{fig:rho_v_fit_v}, and are compared
to those of the FGKPS void model. In this case, one sees that the
$\Delta$ profiles of the two void models are much closer than what was
achieved in the previous subsection, although this comes at the cost
of poorer agreement between the density profiles. In particular, the
spatial extent of the density profile for our void is larger than in
the previous case. 

\begin{figure}
\centering
    \includegraphics[width=\columnwidth]{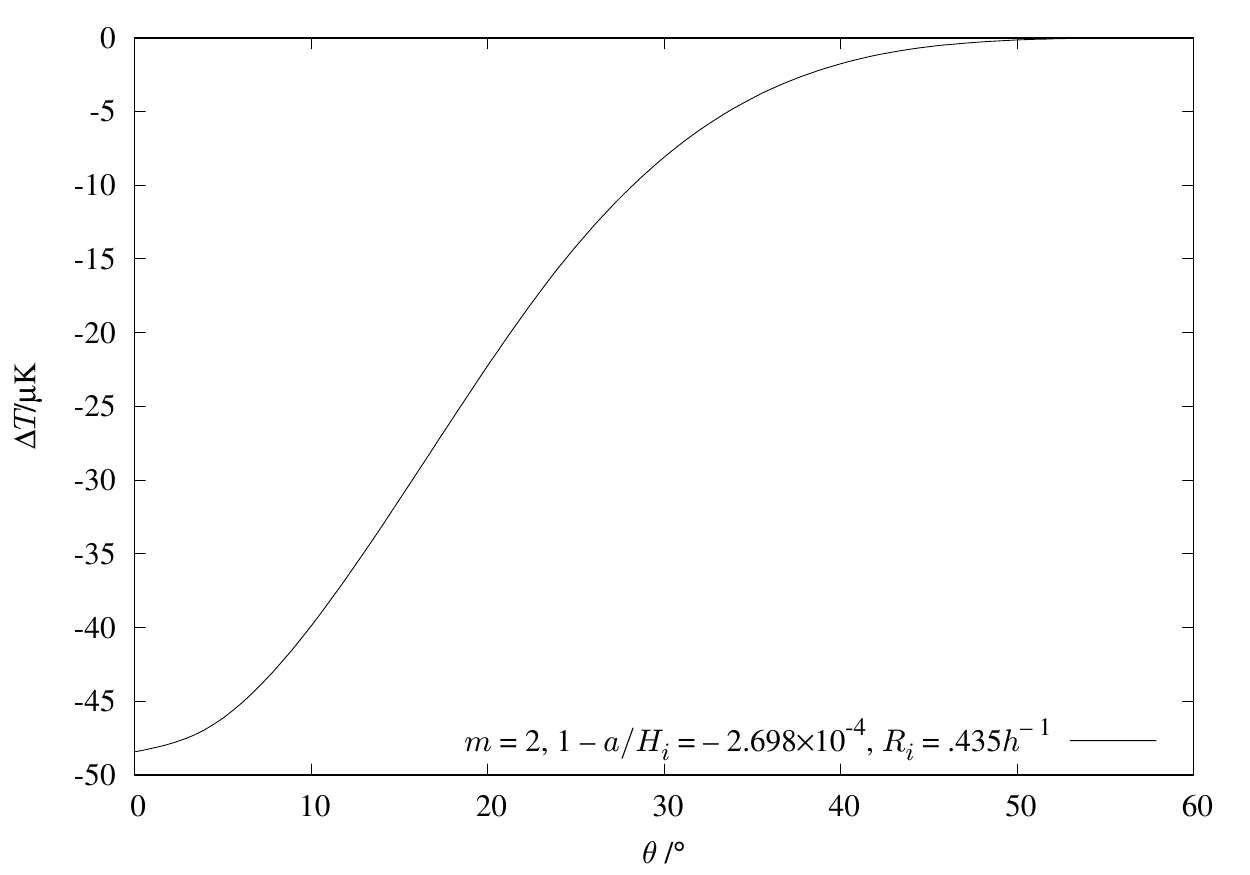}
    \caption{As in Fig.~\ref{fig:dT_plot_fit_void}, but for our void
      model that best reproduces the present-day velocity distribution
      of the FGKPS void.}
    \label{fig:dTabsplot_draco_fitv}
\end{figure}

The larger spatial extent of our void model in this case
requires us to take care in determining the CMB temperature
decrement it produces, since placing the centre of the void at a
redshift $z_c=0.15$ from the observer is insufficient for the observer
to reside in the external universe, beyond the finite compensation
radius of the void. Nonetheless, this may be achieved by placing the
centre of the void at a redshift of $z_c = 0.165$, which corresponds
to the observer lying at a radial coordinate distance of $r=476h^{-1}$
($=600$)~Mpc. The corresponding CMB temperature decrement is plotted
in Fig.~\ref{fig:dTabsplot_draco_fitv}. As expected, the angular
profile of the decrement is larger than that shown in
Fig.~\ref{fig:dT_plot_fit_void} for our previous void model. Most
notable, however, is the central value of $\Delta T \approx
-48$~$\mu$K, which is almost twice that obtained
previously.


\subsection{Reproducing the density and velocity profiles}
\label{sec:fitrhov}

The next obvious approach to consider is to choose the initial
velocity perturbation parameters $R_i$, $a$ and $m$ such that the
resulting present-day density {\em and} velocity profiles of our void
model match those of the FGKPS void model as closely as possible.
As before, this is achieved by performing a numerical optimisation
similar to that used above. In this case, however, we employ a chi-squared
approach in which the uncertainties on the mismatch are taken to be
$\sigma_{\delta} = 0.01$ and $\sigma_{v} = 30 \mathrm{km\,s^{-1}}$ for
the density and velocity profiles, respectively, which ensures that the contribution to the best fit value of $\chi^2$ in roughly similar for the density and velocity profiles. The value of $m$ is fixed to be $2$.

\begin{figure}
\centering
    \includegraphics[width=\columnwidth]{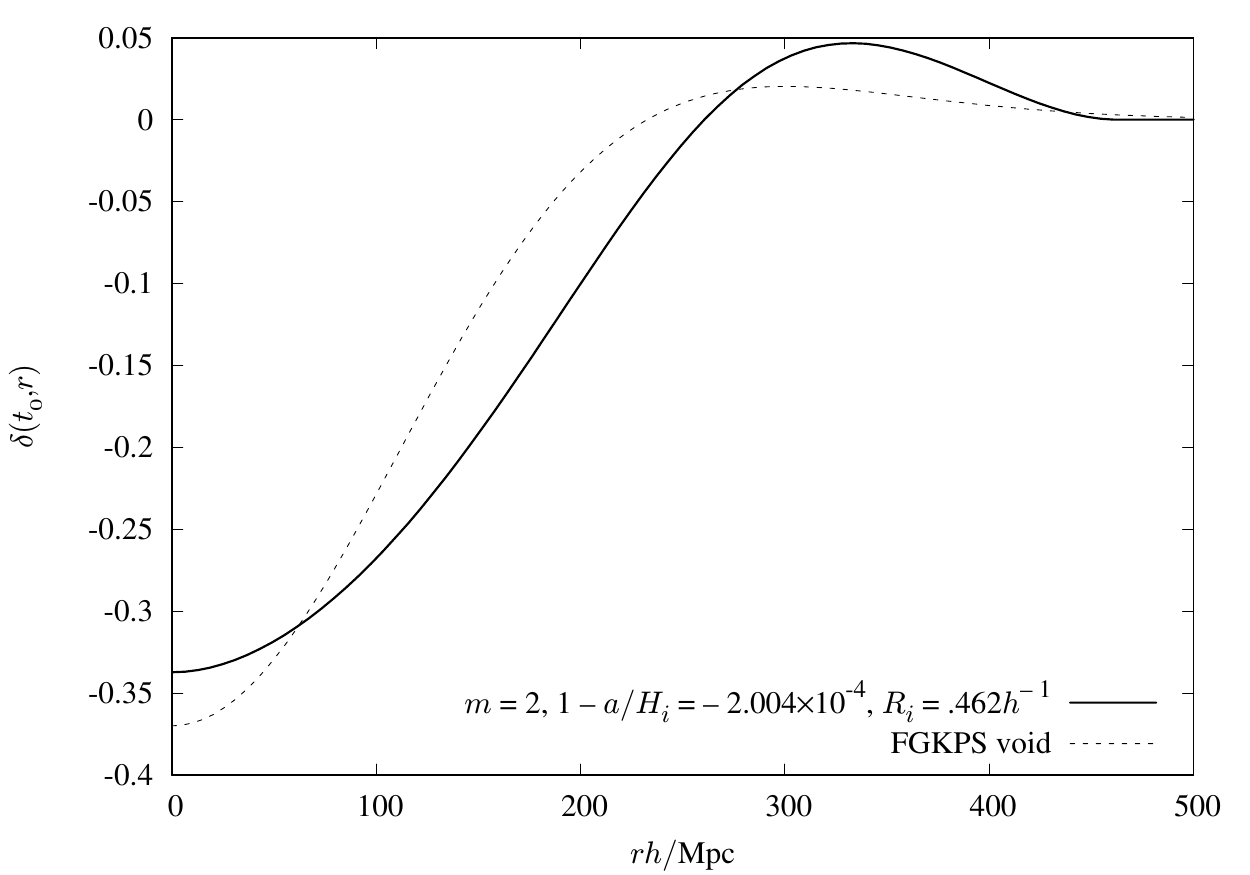} \quad
    \includegraphics[width=\columnwidth]{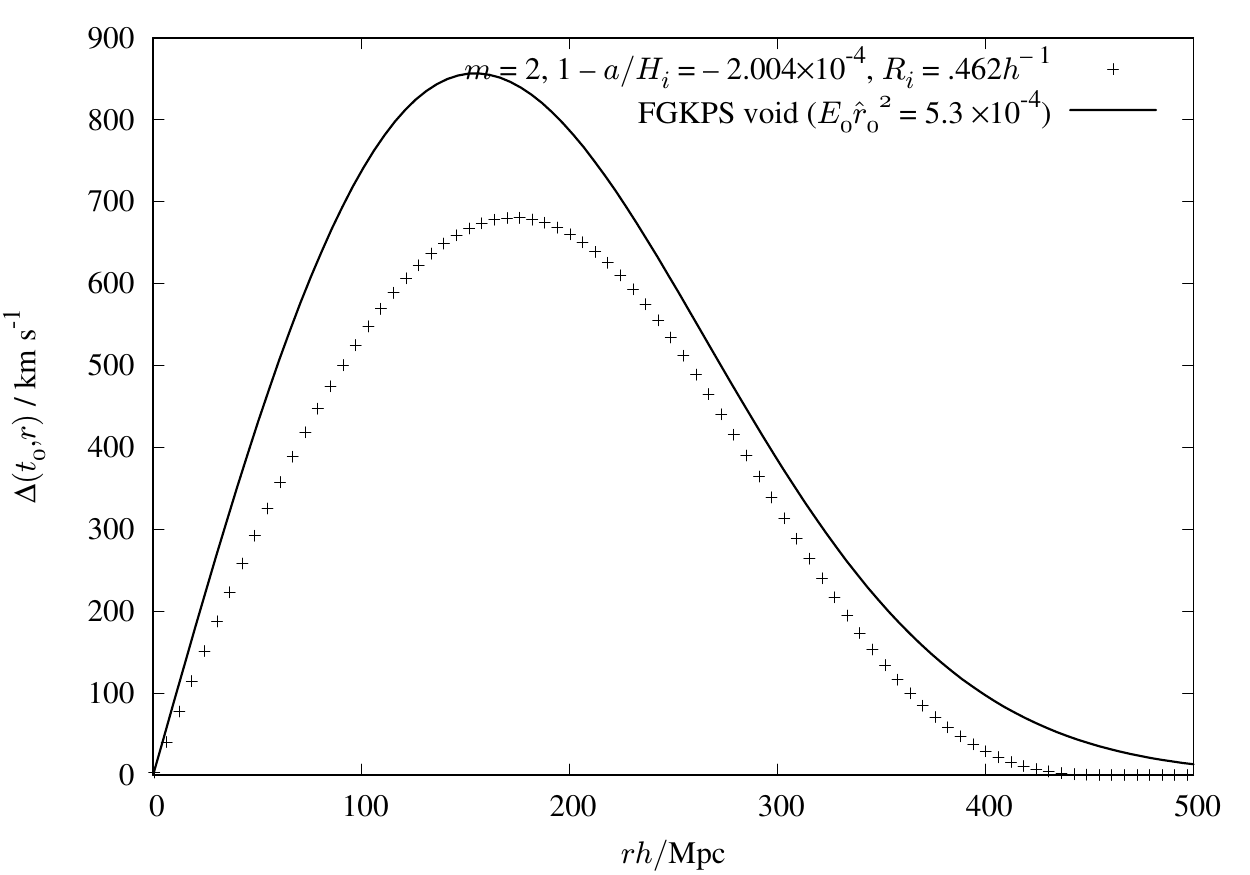}
 \caption{As in Fig.~\ref{fig:rho_v_fit_v}, but for an initial
   velocity perturbation at $z=10^3$ with parameters chosen to
   reproduce the FGKPS void density and velocity profile (solid lines)
   as closely as possible.}
    \label{fig:rho_compare_draco_fitboth2}
\end{figure}

The resulting present-day density and velocity difference $\Delta$
profiles are plotted in Fig.~\ref{fig:rho_compare_draco_fitboth2}, and
are compared to those of the FGKPS void model. As one might expect,
both profiles in our void agree reasonably well with those in the
FGKPS void, but neither agrees as closely as in the case where the
optimisation was performed for that profile alone.  In particular, the
spatial extent of the density profile is similar to that found when
optimising the velocity profile alone, and somewhat larger than when
optimising for the density profile alone. 

\begin{figure}
\centering
    \includegraphics[width=\columnwidth]{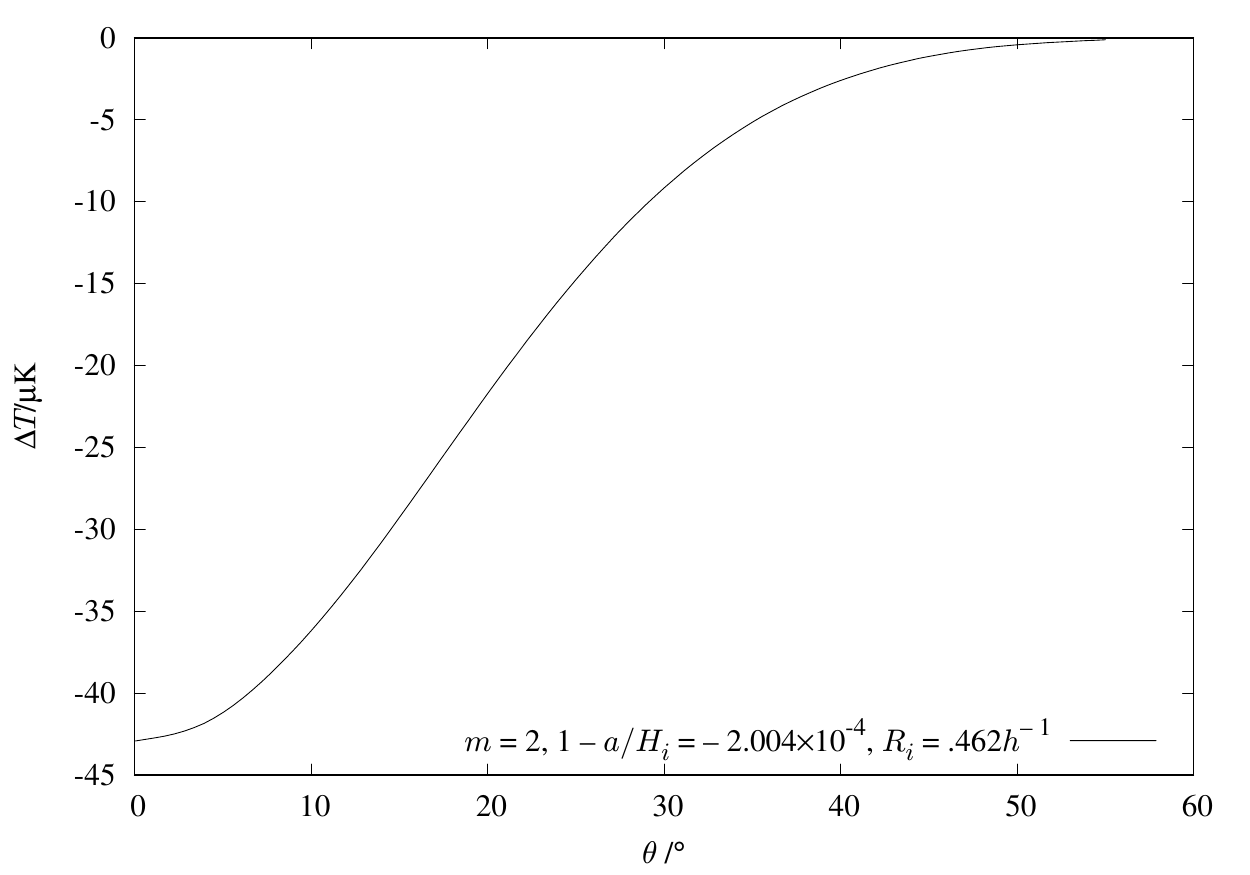}
    \caption{As in Fig.~\ref{fig:dT_plot_fit_void}, but for our void
      model that best reproduces the present-day density and velocity
      distributions of the FGKPS void.}
    \label{fig:dTabsplot_draco_fitboth2}
\end{figure}

As in the previous subsection, the larger spatial extent of our void
model requires us to place the centre of the void at a redshift
$z_c=0.165$ from the observer, in order for the observer to be in the
external universe. The resulting CMB temperature decrement is shown in
Fig.~\ref{fig:dTabsplot_draco_fitboth2}, which is similar in both
angular extent and depth to that plotted in
Fig.~\ref{fig:dTabsplot_draco_fitv}.

\subsection{Consistency with observations}

\begin{figure}
\centering
    \includegraphics[width=\columnwidth]{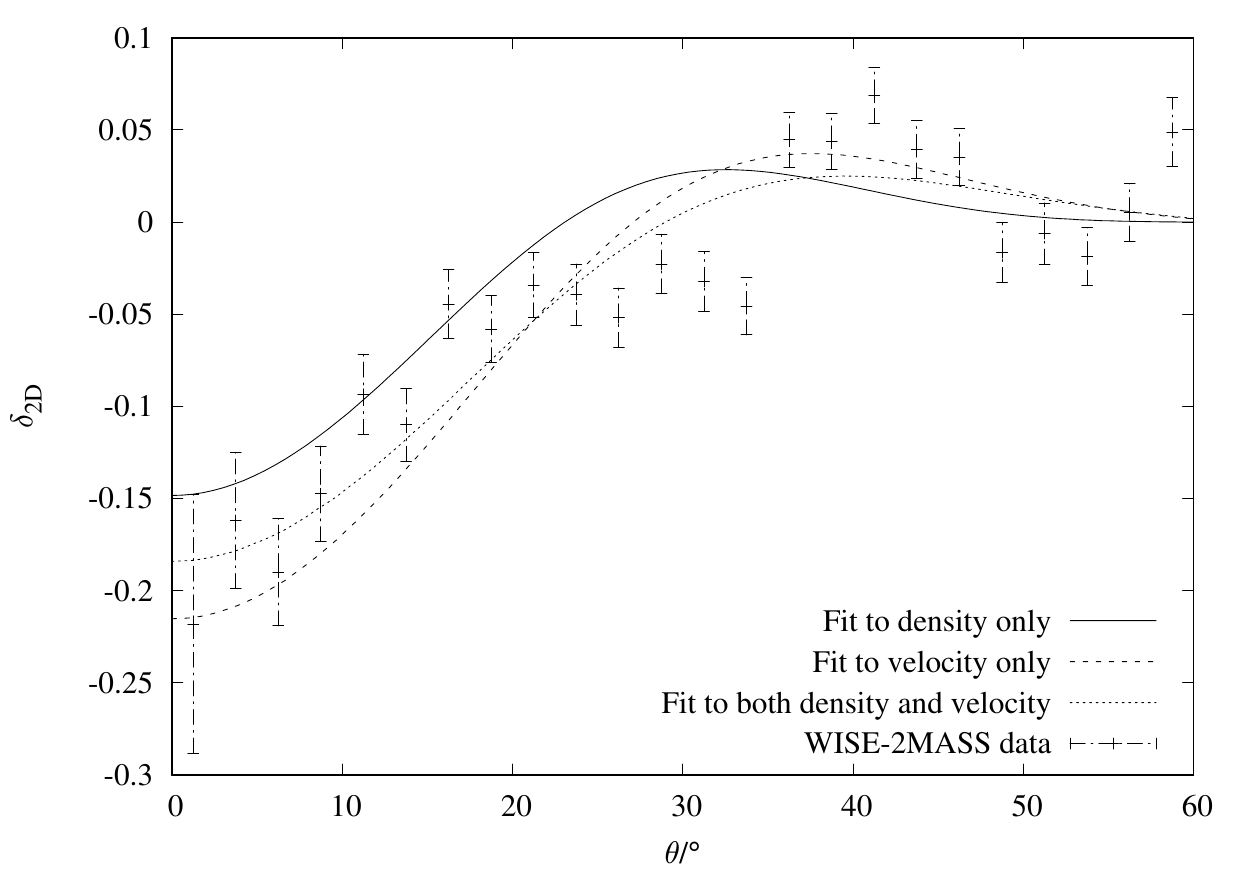}
    \includegraphics[width=\columnwidth]{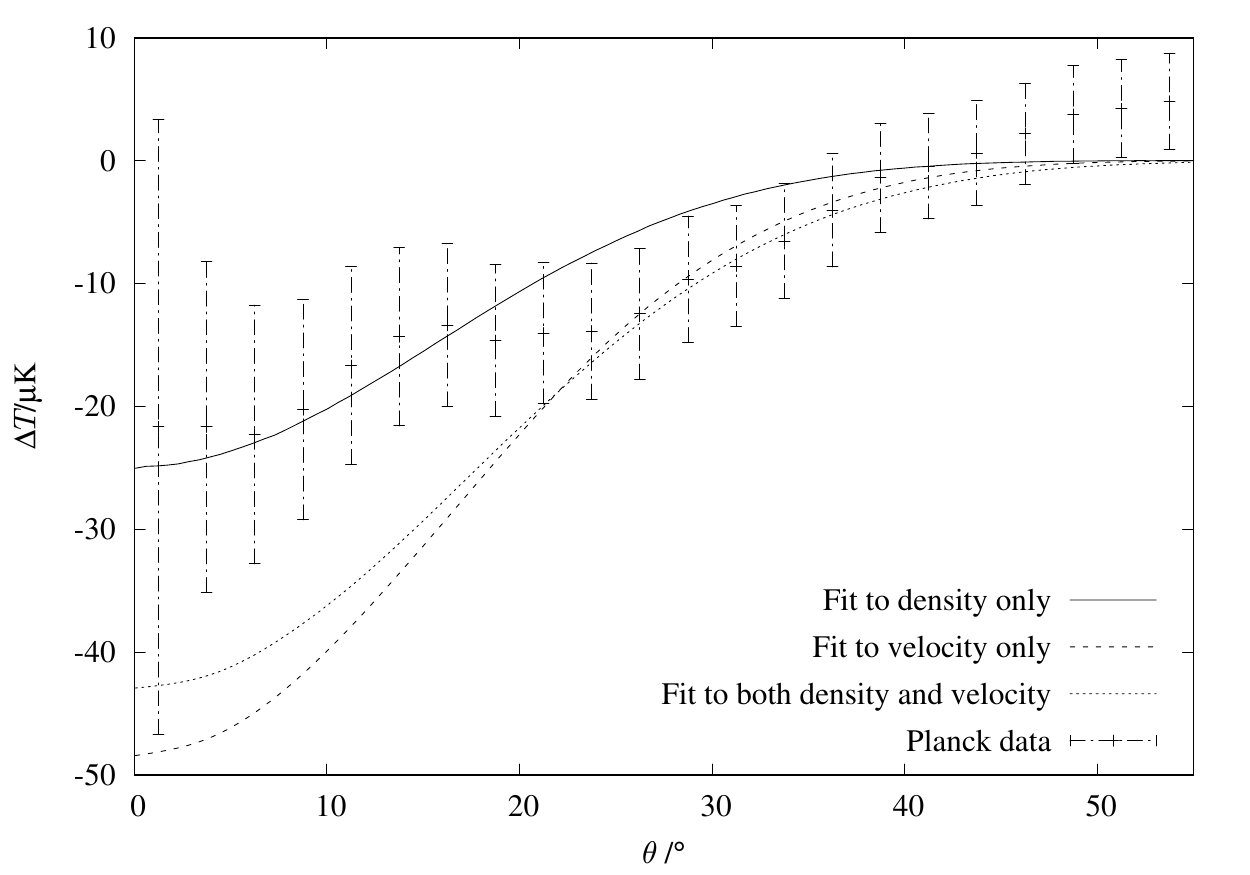} \quad
    \caption{Void projected density profiles (top) and corresponding CMB temperature decrements (bottom) for our three models, compared to data from WISE-2MASS and Planck respectively. Data points are taken from FGKPS.}
    \label{fig:draco_compare_data}
\end{figure}

Although we have focussed on comparing the different characteristics
of the Draco void models produced by the LTB and tetrad-based
approaches, respectively, it is important to determine whether the set of
void models produced using our alternative methodology is consistent
with observations. In Fig~\ref{fig:draco_compare_data}, we thus compare the projected
density profiles and CMB temperature decrements of our three void
models considered in subsections~\ref{sec:fitrho}--\ref{sec:fitrhov} with the WISE-2MASS
galaxy catalogue and the Planck CMB data. The data points are taken
directly from FGKPS. We see that all three void models are
consistent with the WISE-2MASS data. The CMB temperature decrement of our first void model is consistent with the Planck data on all scales, but our second and third void models are consistent with the Planck data only on larger scales, and yield too large a decrement on angular scales below $\sim 20^\circ$.

Although the LTB void model of FGKPS and our three void models are
each consistent with the galaxy counts and broadly consistent with CMB observations in the
direction of Draco (at least on angular scales $\geq 20^\circ$), it is clear that there remain considerable
differences between the models, which result from the different
parameterisations that they employ. Indeed, the above
investigations show that relatively slight differences in the
parameterisation can lead to large changes in the relationship between
density and velocity profiles, and very different CMB temperature
decrements. Moreover, all of the models considered are
consistent with having grown from primordial perturbations in the
early universe. One must therefore be careful in drawing conclusions
regarding the physical characteristics of voids from data that
constrain just their project density distribution and CMB temperature
decrement.

\section{De-evolving the Draco void}
\label{sec:mimic2}

In developing the void models investigated thus far, we have been
careful to ensure consistency with the void having grown from a
primordial perturbation in the early universe. This requirement is, of
course, crucial in producing a physically realistic void model.
Nonetheless, in this section, we investigate the consequences of
neglecting this requirement, with the purpose of demonstrating how
tailoring a void to have given characteristics at the present epoch
results, in general, in a physically unacceptable model.

To this end, we therefore begin by choosing the parameters $R_i$, $a$
and $m$ in our velocity perturbation at $z=10^3$ to produce a velocity
profile today that is as close as possible to that of the FGKPS void
model. We then choose {\em different} values, say $a_{\rho}$ and
$m_{\rho}$, for these parameters, again at $z=10^3$, to produce a
density profile that mimics that of the FGKPS void as closely as
possible. We keep the value of $R_i$ the same in each case, so that
the size of the perturbed region is consistent. We also use the same
value of $H_i$, which ensures that the density profile remains
compensated.

\begin{figure}
\centering
    \includegraphics[width=\columnwidth]{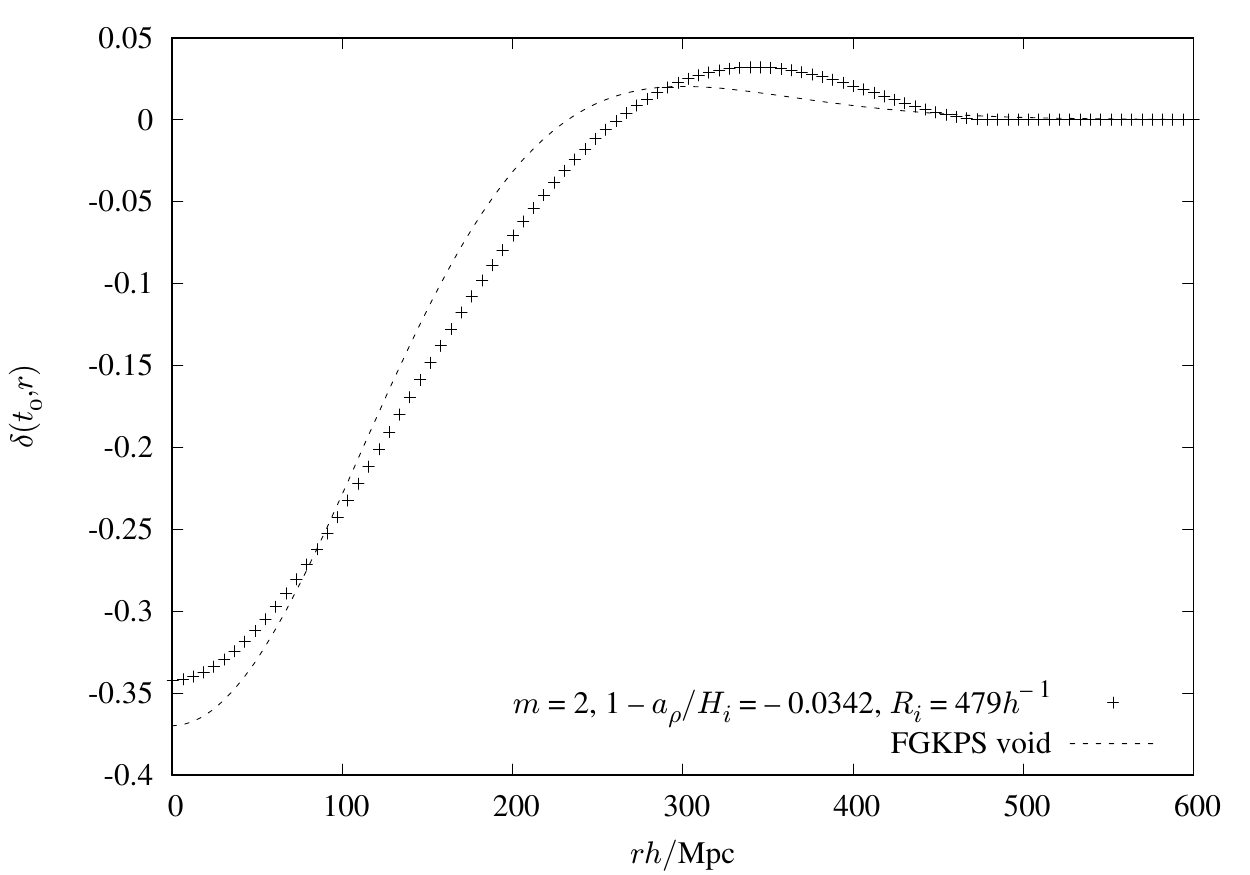} \quad
    \includegraphics[width=\columnwidth]{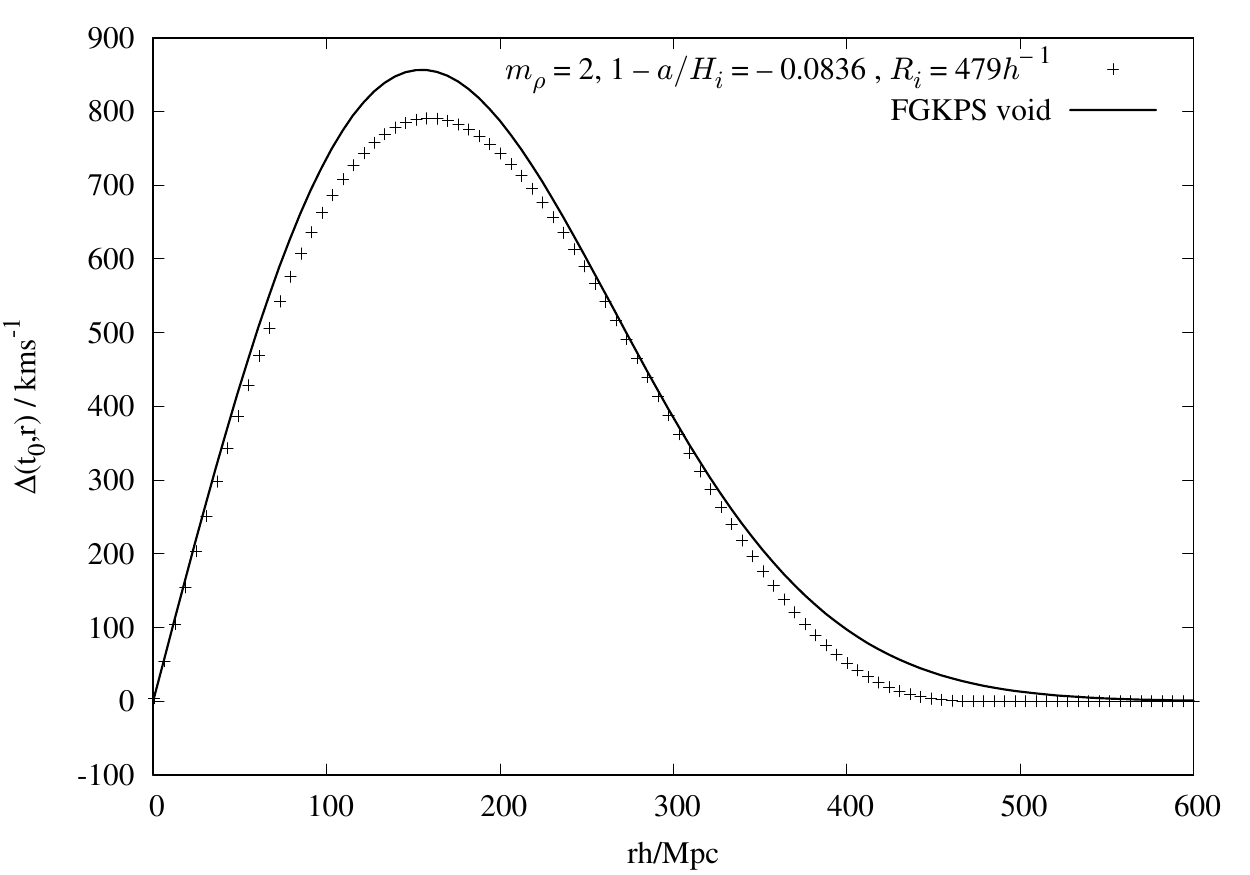}
 \caption{As in Fig.~\ref{fig:rho_compare_draco_fitboth2}, but for two
   initial velocity perturbations at $z=10^3$ chosen separately to
   reproduce the FGKPS void density and velocity profile (solid
   lines), respectively, as closely as possible, and without requiring
   the void to be consistent with having grown from a primordial
   perturbation.}
\label{fig:rho_v_void_now}
\end{figure}

The resulting density and velocity profiles at the current epoch are
shown in Fig.~\ref{fig:rho_v_void_now}. The background Hubble
parameter and the spatial extent of the perturbation are taken as
$H_i=70 \; \mathrm{km\,s^{-1}Mpc^{-1}}$ and $R_i=479 h^{-1}$~Mpc,
respectively. The remaining parameter values that best reproduce the
FGKPS void density profile are $1 - a_{\rho}/H_i=-0.0342$ and
$m_{\rho}=2$, whereas those that best reproduce the FGKPS void
velocity profile (bottom) are $1 - a/H_i=-0.0836$ and $m=2$.

\begin{figure}
\centering
    \includegraphics[width=\columnwidth]{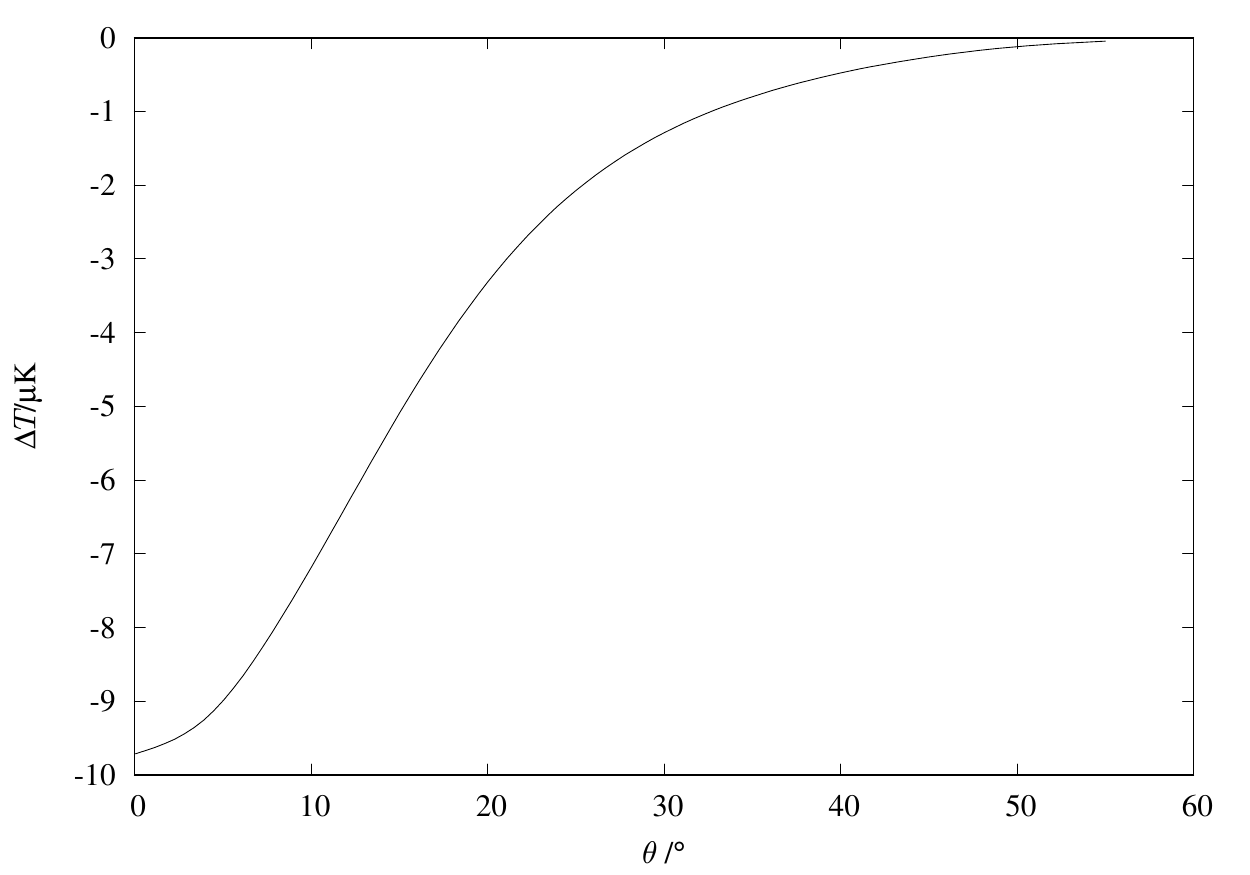}
    \caption{CMB temperature decrement produced by the void with the
      present-day density and velocity profiles plotted in
      Fig.~\ref{fig:rho_v_void_now}.}
    \label{fig:dTabsplot_draco_now_model1}
\end{figure}

To determine the resulting CMB temperature decrement, the large
spatial extent of the void means that one must again place the centre
of the void at a redshift larger than $z_c=0.15$ from the observer for
the latter to reside in the external universe. We find that this can
be achieved by placing the observer at a radial coordinate distance of
$483 h^{-1}$ Mpc, which corresponds to $z_c \sim 0.167$.  The
resulting CMB temperature decrement is plotted in
Fig.~\ref{fig:dTabsplot_draco_now_model1}. Interestingly,
the depth of the decrement is much smaller than found for our previous
void models, with a central value of just $\Delta T \approx -9.7$ $\mu$K.

\begin{figure*}
\centering
\includegraphics[width=2\columnwidth]{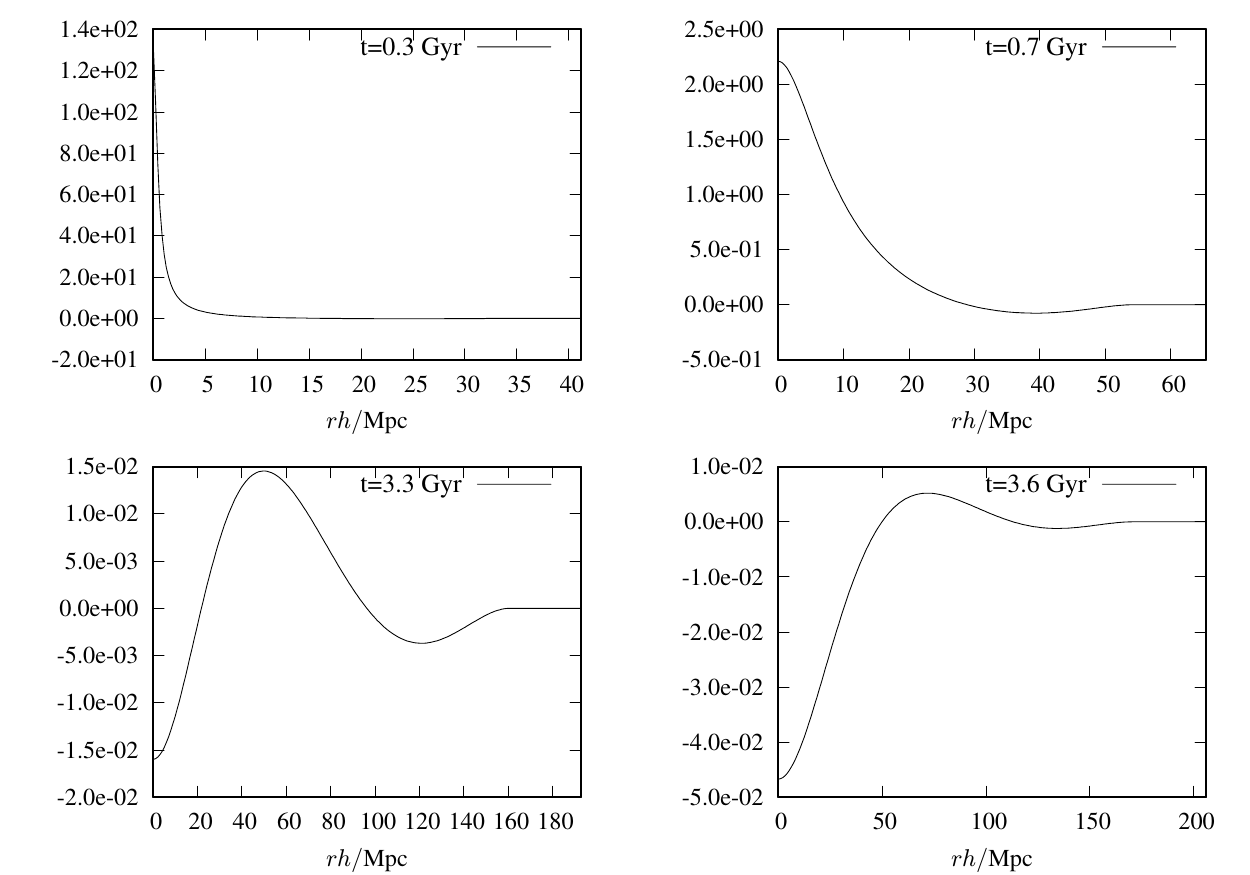}
\caption{The density contrast at a selection of earlier epochs for the
  void with the present-day ($t=13.5$~Gyr) density and velocity profiles
  plotted in Fig.~\ref{fig:rho_v_void_now}.}
    \label{fig:rho_draco_M1_evolve}
\end{figure*}

One may demonstrate that this void model is physically unrealistic,
however, by considering the void with the present-day density and
velocity profiles plotted in Fig.~\ref{fig:rho_v_void_now} and
evolving it {\em backwards} in time. The value of the $t$ parameter
corresponding to the current epoch is 13.5~Gyr. The density
contrast at a selection of earlier times is shown in
Fig.~\ref{fig:rho_draco_M1_evolve}. One sees that the density contrast
diverges at early epochs. This occurs because the void model
contains a decaying mode, which grows as one moves backwards in time.
Consequently, this void model is not consistent with having grown from
a primordial perturbation. This behaviour is, in fact, quite generic
unless one takes care to exclude the decaying mode by choosing the initial
density and velocity profiles in the early universe to obey the
condition \eqref{eqn:ic_rho_flat_L}.

\section{Conclusions}
\label{sec:conc}

We apply our tetrad-based approach for constructing
spherically-symmetric solutions in general relativity to modelling
voids and the secondary anisotropies that they induce in the CMB.  We
compare our approach to the usual LTB method, and demonstrate that the
two methods represent a Eulerian and Lagrangian description,
respectively, of the dynamics of a pressureless cosmological fluid.

In particular, we use our approach to construct models for the void
observed in the direction of Draco in the WISE-2MASS galaxy survey,
and a corresponding CMB temperature decrement in the Planck data in
the same direction, and compare our void models with that produced by
\cite{Finelli2016} using the LTB formalism. We find that the
present-day characteristics of the void, summarised by its current
density and velocity profiles, are not well constrained by the
existing data, such that a large range of different void models are
broadly consistent with the observations. In particular, we note that models
derived from different parameterisations of the void typically lead to
very different density and/or velocity profiles. CMB temperature decrements are especially sensitive to the velocity profile; however it is often overlooked in LTB models of voids.

Finally, we demonstrate the importance of ensuring that void models
are consistent with having evolved from primordial perturbations in
the early universe, and hence contain no contribution from a decaying
mode. In particular, we show that constructing a void model such that
it has given density and velocity profiles at the present epoch will,
in general, lead to unphysical singularities in the void model at
earlier epochs.



\section*{Acknowledgements}
Do Young Kim is supported by a Samsung Scholarship.  We thank Seshadri
Nadathur for useful discussions related to over-constraint of an LTB model.








%
%
%
%
%
\end{document}